\documentclass[aps,showpacs,prl,twocolumn]{revtex4-2}

\usepackage[pdftex]{graphicx}
\usepackage{amssymb}
\usepackage{amsmath}
\usepackage{xspace}
\usepackage{color}
\usepackage{hyperref}

\usepackage[resetlabels]{multibib}
\newcites{SM}{Refs in Supplemental Material}

\newcommand{\tH}{t_{\rm h}}

\begin{document} 

\title{Radiative Higgs mode in photoinduced $\eta$ pairs}

\author{Satoshi Ejima$^{1,2}$}
\author{Benedikt Fauseweh$^{3,4}$}
\affiliation{
$^1$ Institute of Software Technology, German Aerospace Center (DLR), 22529 Hamburg, Germany\\
$^2$ Computational Condensed Matter Physics Laboratory,
     RIKEN Cluster for Pioneering Research (CPR), Saitama 351-0198, Japan\\
$^3$ Institute of Software Technology, German Aerospace Center (DLR), 51147 Cologne, Germany\\
$^4$ Department of Physics, TU Dortmund University, Otto-Hahn-Str.~4, 44227 Dortmund, Germany
}

\date{\today}

\begin{abstract}
  We demonstrate that the $\eta$ symmetry of the pump-induced $\eta$-pairing state in the Hubbard chain can be transiently broken by an additionally applied probe pulse. This leads to characteristic dynamic Higgs oscillations of the $\eta$-pair correlations during the probe pulse and a sharp dynamic negative optical conductivity $\sigma(\omega;t)$, which appears above the equilibrium Mott gap $\omega>\Delta_{\rm c}$, long after the pump pulse is finished. This negative peak in the real part of $\sigma(\omega;t)$ is a distinctive hallmark of the photoinduced $\eta$-pairing state, distinguishing it from parameter regimes dominated by incoherent doublon formation, which do not exhibit this feature. Remarkably, the broadband probe pulse then leads to the emission of photons with the Higgs frequency.
\end{abstract}

\maketitle

Recent developments in laser technology have revolutionized various scientific fields, enabling exceptional precision and control. Innovations such as ultrafast pulse lasers have opened new frontiers in quantum physics and material science, and have also found significant applications in condensed matter physics, particularly in the study of strongly correlated electron systems~\cite{ Giannetti2016,Ishihara2019,RevModPhys.93.041002}. Examples include the ultrafast switching of Weyl semimetals \cite{Sie2019}, anomalous Hall effect in graphene~\cite{McIver2019} and charge-density waves in rare-earth tritelluride LaTe$_3$~\cite{Kogar2019}. 

A particularly interesting prospect is the use of femtosecond laser pulses
to dynamically induce novel states in quantum materials~\cite{Basov2017}, with light-induced superconductivity being the most prominent example ~\cite{Mankowsky2014,Mitrano2016,PhysRevX.10.031028,Budden2021}. Significant theoretical efforts have been devoted to understanding the conditions under which such exotic non-equilibrium states emerge ~\cite{PhysRevB.81.115131, doi:10.7566/JPSJ.88.044704, PhysRevB.96.054506, PhysRevB.101.180507, 10.21468/SciPostPhys.15.6.236}, yet the mechanisms and detection of these phases remain under debate.

In this context, one intriguing phenomenon that has attracted attention is the $\eta$ pairing, first proposed by Yang for the Hubbard model~\cite{PhysRevLett.63.2144}, which represents a state with off-diagonal long-range order that could be important for understanding high-temperature superconductivity and other quantum phenomena. 
However, $\eta$ pairs are absent in the ground state and have received only limited attention, especially from a theoretical point of view. A recent study has shown that pulse irradiation can induce $\eta$ pairing in the Hubbard model, even in its Mott insulating phase~\cite{Kaneko19}. The nonlinear optical response is crucial in enhancing the number of $\eta$ pairs, thereby promoting superconducting correlations in the photoexcited state. These findings highlight the potential of nonequilibrium dynamics as a way to access exotic quantum states and enhance superconductivity \cite{PhysRevLett.125.137001}. The synergy between advanced laser technologies and the Hubbard model provides a promising avenue for exploring new quantum phases and understanding the mechanisms underlying high-temperature superconductivity, as well as the behavior of cuprate superconductors dominated by one-dimensional (1D) physics, such as  Ba$_{2-x}$Sr$_x$CuO$_{3+\delta}$ and Pr$_2$Ba$_4$Cu$_7$O$_{15-\delta}$.~\cite{science.373.1235,MATSUKAWA2004101,JPSJ.80.024708}.

The detection of $\eta$ pairs in experimental settings presents significant challenges due to the intricate nature of these quantum states. First, the Hamiltonian of the system must preserve the number of $\eta$ pairs, corresponding to a hidden SU(2) symmetry. This requirement is not easily met in typical experimental setups, making it difficult to create and maintain the required conditions for $\eta$ pairs. In addition, the transient nature of photoinduced $\eta$ pairs adds another layer of complexity. These pairs are typically short-lived and require ultrafast measurement techniques to capture their dynamics. The use of advanced laser systems, such as ultrashort pulse lasers, is essential to induce and probe these states, but even with these tools, the signal can be weak and easily obscured by noise.

As a candidate physical quantity, Kaneko {\it et al}. demonstrated one of the hallmarks of superconductivity, the nonvanishing charge stiffness of the photoinduced  $\eta$-pairing states, by analyzing the system-size dependence using the exact diagonalization method~\cite{Kaneko20}.
Time- and angle-resolved photoemission spectroscopy experiments would provide clear signatures of the photoinduced insulator-to-metal quantum phase transition~\cite{Ejima22}. Although performing such experiments in optical lattices remains highly challenging, they are neverthless strong candidates for realizing Hubbard-type models. The other candidate might be the entanglement entropy, showing the entanglement growth due to the photoinduced $\eta$-pairing state~\cite{Ejima23}, since the von Neumann entropy and the second-order R\'{e}nyi entropy can be detected in optical lattices~\cite{Islam2015,Kaufman2016, PhysRevLett.101.265301}. 

Here, we demonstrate that the nonequilibrium optical conductivity $\sigma(\omega; t)$ shows a characteristic negative spectral weight only for the optimal pump-pulse parameter set, which maximally enhances the photoinduced $\eta$-pairing state. The integrated negative spectral weights $f(t)$ coincide almost perfectly with those from the pair correlations after the Fourier transformation. The time dependence of $f(t)$ shows coherent oscillations similar to previously reported Higgs oscillations in conventional $s$-wave and unconventional $d$-wave superconductors \cite{annurev:/content/journals/10.1146/annurev-conmatphys-031119-050813}. 

\begin{figure}[tb]
    \includegraphics[width=0.99\linewidth]{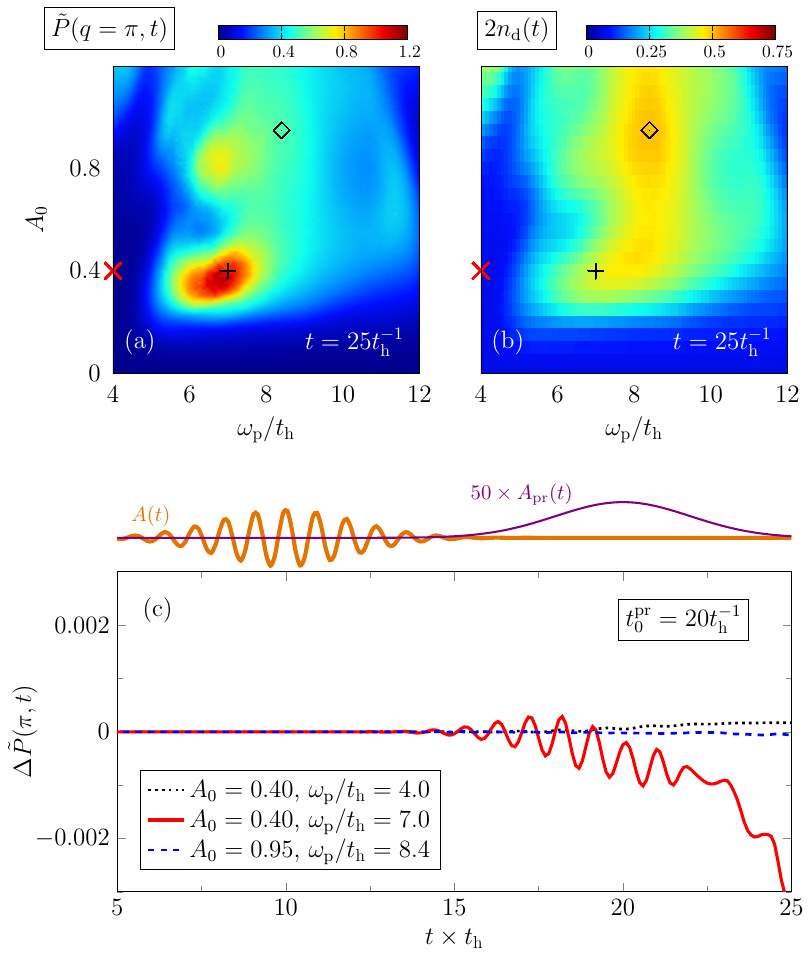}
    \caption{Contour plots of $\tilde{P}(q=\pi,t)$ (a) and $2n_{\rm d}(t)$ (b) at $t=25\tH^{-1}$ in the $\omega_{\rm p}$-$A_0$ plane 
    for the half-filled Hubbard chain with $U/\tH=8$, obtained by iTEBD.
    See also Ref.~\cite{Ejima22} at $t=15\tH^{-1}$. $\sigma(\omega;t)$ will be demonstrated at the symbols `+' and `$\diamond$' where $\eta$-pairing-dominant ($A_0=0.4$ and $\omega_{\rm p}/\tH=7.0$) and doublon-dominant ($A_0=0.95$ and $\omega_{\rm p}/\tH=8.4$)  points, respectively, contrasting with the non-dominant ($A_0=0.4$ and $\omega_{\rm p}/\tH=4.0$) point (`$\times$') at $t=25\tH^{-1}$. (c): The deviation of $\eta$-pair correlations $\Delta\tilde{P}(\pi;t)$ at characteristic three points denoted by the symbols in panels (a) and (b), exhibiting the radiative Higgs oscillations at the $\eta$-pairing dominated point (`+').
    }
    \label{PpitNd}
  \end{figure}
 
 The Hamiltonian of the 1D half-filled Hubbard model is defined as
 \begin{align}
  \hat{H}=&
  -\tH \sum_{j,\sigma}
   \big(
    \hat{c}_{j,\sigma}^\dagger \hat{c}_{j+1,\sigma}^{\phantom{\dagger}}
    +\text{H.c.}
   \big)
   \nonumber \\
  &
   +U\sum_{j}\left(
      \hat{n}_{j,\uparrow}-1/2 
     \right)
             \left(
     \hat{n}_{j,\downarrow}-1/2 
    \right)\,,
   \label{hubbard}
 \end{align}
 where $\hat{c}_{j,\sigma}^{\dagger}$ ($\hat{c}_{j,\sigma}^{\phantom{\dagger}}$)
 is the creation (annihilation) operator of an electron with spin projection 
 $\sigma\in\{\uparrow,\downarrow\}$ at lattice site $j$, and 
 $\hat{n}_{j,\sigma}=\hat{c}_{j,\sigma}^{\dagger} \hat{c}_{j,\sigma}^{\phantom{\dagger}}$ 
 is the number operator. $\tH$ and $U$ are nearest-neighbor hopping amplitude and
 on-site Coulomb repulsion ($U>0$), respectively.
 The so-called $\eta$ operators, introduced in the seminal paper by Yang~\cite{PhysRevLett.63.2144},
 construct exact eigenstates of the Hubbard model, 
 \begin{align}
  \hat{\eta}^+ &= \sum_j(-1)^j \hat{\Delta}_j^\dagger
  \equiv \sum_j \hat{\eta}_j^+\, ,
  \ \  \ 
  \hat{\eta}^- = (\hat{\eta}^+)^\dagger\, ,
  \label{etapm}
 \\
  \hat{\eta}^z &= \frac{1}{2}\sum_{j}  (\hat{n}_{j,\uparrow}+\hat{n}_{j,\downarrow}-1)
  \equiv \sum_j \hat{\eta}_j^z \, ,
  \label{etaz}
 \end{align}
 which obey the SU(2) commutation relations.
 Here, $\hat{\Delta}_j^\dagger=\hat{c}_{j,\downarrow}^\dagger\hat{c}_{j,\uparrow}^\dagger$ is 
 the on-site singlet-pair creation operator. 
 Since the Hubbard Hamiltonian~\eqref{hubbard} commutes with the operator  
 $\hat{\eta}^2=\tfrac{1}{2}(\hat{\eta}^+\hat{\eta}^-+\hat{\eta}^-\hat{\eta}^+) +  (\hat{\eta}^z)^2$, 
 so that Hubbard eigenstates are also eigenstates of $\eta^2$, and most importantly 
 eigenstates with a finite value of $\langle \hat{\eta}^2 \rangle$ have long-ranged pairing correlations 
 $\langle\hat{\eta}^+_j\hat{\eta}^-_{\ell}\rangle$~\cite{PhysRevLett.63.2144}.

 As demonstrated in Ref.~\cite{Kaneko19}, $\eta$ pairs can be induced by applying 
 a pump pulse to Mott insulators described theoretically 
 with the gauge transformation 
 $\tH \hat{c}_{j,\sigma}^\dagger \hat{c}_{j+1,\sigma}^{\phantom{\dagger}}  
 \rightarrow \tH e^{\mathrm{i}A(t)} \hat{c}_{j,\sigma}^\dagger
                        \hat{c}_{j+1,\sigma}^{\phantom{\dagger}} 
 $,
 where $A(t)$ is the vector potential associated with the external electric field 
 of the pump pulse
 \begin{eqnarray}
   A(t)=A_0 e^{-(t-t_0)^2/(2\sigma_{\rm{p}}^2)}\cos\left[\omega_{\rm{p}}(t-t_0)\right]\,,
  \label{pump}
 \end{eqnarray}
 where $A_0$ is the amplitude, $\omega_{\rm p}$ is the frequency and
 $\sigma_{\rm p}$ is the width centered at time $t_0$ ($>0$). 
 This so-called Peierls substitution makes the Hamiltonian time-dependent
 $\hat{H}\to\hat{H}(t)$. By utilizing the infinite time-evolved block decimation 
 (iTEBD) technique~\cite{iTEBD} with second-order Suzuki--Trotter 
 decomposition, the initial ground state evolves in time as 
 $|\psi(0)\rangle\to |\psi(t)\rangle$.
 In the following, we take $\tH$ ($\tH^{-1}$) as the unit of energy (time) 
 and set the time step $\delta t=0.01\tH^{-1}$.
 In this study we use the pump pulse with width $\sigma_{\rm p}=2\tH^{-1}$ centered at time $t_0=10\tH^{-1}$. See also the Supplemental Material~\cite{SM} for further technical details and the choice of pump-pulse parameters.

 Note that in our system a relaxation processes of the photoinduced $\eta$ pairs are not included, such as the phonon degrees of freedom. We expect that in experiment the photoexcited $\eta$ pairing state can be observed shortly after the pump-pulse irradiation, i.e., within a few hundred femtoseconds.

\begin{figure*}[tb]
  \includegraphics[width=0.95\linewidth]{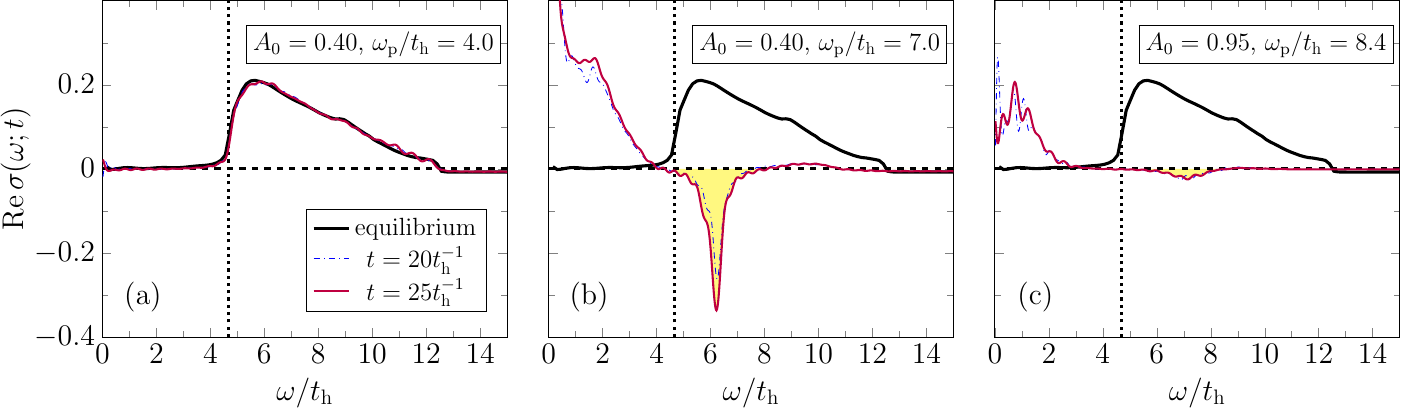}
  \caption{Nonequilibrium optical conductivity $\sigma(\omega; t)$ for 
  non-dominant (`$\times$' symbol in Fig.~\ref{PpitNd}), $\eta$-pairing-dominant (`+') and doublon-dominant ('$\diamond$')
  pump-pulse parameters from left to right panels.
  The black line is the optical conductivity $\sigma(\omega)$ at equilibrium. The blue (red) line exhibits $\sigma(\omega; t)$ at $t=20\tH^{-1}$ ($t=25\tH^{-1}$), respectively. 
  The vertical dotted line denotes the position of the Mott gap, $\omega=\Delta_{\rm c}$. 
  }
  \label{sigma-w}
\end{figure*}

To detect the photoinduced $\eta$-pairing state, we simulate the time evolution of the pair correlations
\begin{align}
 P(r,t) =\frac{1}{L}\sum_j \langle\psi(t)|\hat{\Delta}_{j+r}^\dagger\hat{\Delta}_j +{\rm H.c.}|\psi(t)\rangle
\end{align}
and its Fourier transform $\tilde{P}(q,t)=\sum_r e^{\mathrm{i}qr}P(r,t)$, where $L$ is the number of lattice sites. 
Note that the pair correlation at $r=0$ is equal to the number of double occupancy,
$P(0,t)=2n_{\rm d}(t)=(2/L)\sum_j \langle\psi(t)|\hat{n}_{j,\uparrow}\hat{n}_{j,\downarrow} |\psi(t)\rangle$.

Figures~\ref{PpitNd}(a) and (b) demonstrate the $\omega_{\rm p}$ and $A_0$ dependence of $\tilde{P}(q=\pi,t)$
and $2n_{\rm d}(t)$, respectively, for the time $t=25\tH^{-1}$, which is related to the nonequilibrium optical conductivity results discussed later.   
Instead of an artificial stripe structure seen in the previous studies with small clusters~\cite{Kaneko19,SCES19}, the single peak structure appears around $\omega_{\rm p}/\tH=7.0$ and $A_0=0.4$ in Fig.~\ref{PpitNd}(a) by simulating the system directly in the thermodynamic limit ($L\to\infty$) using iTEBD, while the double occupancy is only slightly enhanced around $\omega_{\rm p}/\tH\approx U$ as seen in Fig.~\ref{PpitNd}(b). The longer-range contributions of the $\eta$-pairing correlations play a significant role around the peak region of Fig.~\ref{PpitNd}(a).
In addition to the pump pulse we apply a weak probe pulse 
$A_{\rm pr}(t)=A_0^{\rm pr}e^{-(t-t_0^{\rm pr})^2/2\sigma_{\rm pr}^2}$ 
centered in time around $t_0^{\rm pr}=20\tH^{-1}$, $\sigma_{\rm pr}=2.0 \tH^{-1}$ and zero frequency. 
In Fig.~\ref{PpitNd}(c) we show the difference 
$\Delta\tilde{P}(\pi; t)=\tilde{P}_\mathrm{pump+pr}(\pi; t)-\tilde{P}_\mathrm{pump}(\pi; t)$ between the pump and probe induced $\eta$ correlations.  The probe pulse transiently breaks the $\eta$ symmetry, inducing Higgs oscillations in the $\eta$ correlations. In the non-dominant and doublon-dominant region, no oscillations are observable, consistent with the dependence of the Higgs oscillation on the size of the superconducting order parameter. Note that the pump pulse has decayed up until the probe pulse, and does not induce these oscillations, i.e. they are intrinsic to superconducting condensate. In the following we discuss the consequence of these oscillations in the dynamical response of the system.

Let us now focus on the nonequilibrium optical conductivity.
In the presence of $A(t)$, the current operator $\hat{J}$ also becomes time dependent:
\begin{align}
 \hat{J}(t)=-\frac{\partial \hat{H}(t)}{\partial A(t)}
 =\tH\sum_{j,\sigma}
  \left(\mathrm{i}e^{\mathrm{i}A(t)}
   \hat{c}_{j,\sigma}^\dagger \hat{c}_{j+1,\sigma}^{\phantom{\dagger}}
   +{\rm H.c.}
   \right)\,.
\end{align}
The probe pulse induces a current deviation 
$J_{\rm pr}(t)=\langle\hat{J}_{A+A_{\rm pr}}\rangle_t-\langle\hat{J}_{A}\rangle_t$. To simulate a time resolved experiment, we calculate the optical conductivity using a narrow probe pulse characterized by $A_0^{\rm pr}=0.05$ and $\sigma_{\rm pr}=0.05$ with the delay time between pump and probe pulses $\tau=t_0^{\rm pr}-t_0$. The nonequilibrium optical conductivity is then given by~\cite{PhysRevB.93.195144,PhysRevB.104.085122} 
\begin{align}
\sigma(\omega,\tau) 
 =\frac{j_{\rm pr}(\omega)}{\mathrm{i}(\omega+\mathrm{i}\gamma)A_{\rm pr}(\omega)}\, ,
 \label{eq:sigma-omega}
\end{align}
where $A_{\rm pr}(\omega)$ and $j_{\rm pr}(\omega)$ are the Fourier transformations of 
$A_{\rm pr}(\omega)$ and $j_{\rm pr}(t)[=J_{\rm pr}(t)/L]$, respectively.
The damping factor $\gamma(=0.1)$ is introduced when the Fourier transformations are performed
due to the finite simulation time. This is also necessary to distinguish the Drude component of the spectral weight in the limit $\omega\to0$. In this paper, we rewrite Eq.~\eqref{eq:sigma-omega} as 
$\sigma(\omega;t)$ with redefining $t=t_0^{\rm pr}=t_0+\tau$ to compare the result of pair correlation functions shown in Fig.~\ref{PpitNd}. 

Note that the expectation value of $\hat{J}(t)$ in (quasi-) 1D systems can be simulated directly in the thermodynamic limit by iTEBD, which allows us to observe $\sigma(\omega,t)$ for $L\to\infty$, i.e., in the absence of finite-size and boundary effects. For more details, see Ref.~\cite{Sugimoto23}.

Figure~\ref{sigma-w} shows the optical conductivity $\sigma(\omega;t)$ after pulse irradiation ($t=25\tH^{-1}$) in the $\eta$-pairing nondominant (a) and dominant (b) regimes in addition to the doublon-dominant region (c). 
In panel (a) for $\omega_{\rm p}/\tH=4.0$ and $A_0=0.40$ [marked in Fig.~\ref{PpitNd} as `$\times$'], where the pair correlation $P(r,t)$ doesn't enhance, $\sigma(\omega; t)$ is almost equivalent to the one in equilibrium as expected. Namely, the spectral weight becomes positive finite above the Mott gap 
($\omega\gtrsim \Delta_{\rm c}$ with $\Delta_{\rm c}/\tH\simeq 4.68$ for $U/\tH=8$, see the vertical dotted line) and $\sigma(\omega; t)$ are almost equal to the optical conductivity at equilibrium $\sigma(\omega)$. Moreover, $\sigma(\omega;t)$ exhibits no significant time dependence, consistent with the former results of photoemission spectra at nonequilibrium~\cite{Ejima22}.
The situation changes significantly in the $\eta$-pairing dominant regime as in Fig.~\ref{sigma-w}(b) for $A_0=0.4$ and $\omega_{\rm p}/\tH=7.0$ (marked as `$+$' in Fig.~\ref{PpitNd}). After pump-pulse irradiation, the spectral weight above the Mott gap is not positive at all but negative. Most remarkably, it exhibits a sharp peak structure, which differs from the former results of the Hubbard model in infinite dimensions~\cite{PhysRevB.102.165136,PhysRevB.108.174515}. This sharp peak is the hallmark of a coherent mode that dominates the negative conductivity.
The small enhancement of the pair correlations also occurs around $\omega_{\rm p}\approx U$ due to the doublon formation [Fig.~\ref{PpitNd}(b)]. The double occupancy  $2n_{\rm d}(t)$ is most enhanced around $A_0\approx0.95$ and $\omega_{\rm p}/\tH\approx8.4$ marked as `$\diamond$' in Fig.~\ref{PpitNd}. With this pump-pulse parameter set, $\sigma(\omega; t)$ is also not positive for $\omega_{\rm p}\gtrsim\Delta_{\rm c}$ as in the $\eta$-pairing dominant regime, but the spectral weight is only slightly negative as shown in Fig.~\ref{sigma-w}(c).  
Thus, the negative conductivity with a sharp peak structure can be a fingerprint of the photoinduced $\eta$-pairing state in pump-probe experiments. Importantly, this contradicts previous interpretations of the negative conductivity as coming purely from doublon-hole recombination \cite{PhysRevB.102.165136,PhysRevB.108.174515}. In the following, we demonstrate that the photoinduced $\eta$ pairs can be searched by integrating out the (negative) spectral weight of $\sigma(\omega, t)$ for $\omega\gtrsim \Delta_{\rm c}$. 

Note that in Ref.~\cite{PhysRevB.102.165136} Li {\it et al.} discussed the Drude weight $D$, which is given by ${\rm Re}\, \sigma(\omega)\sim D\delta(\omega)$,  since this relates to the $\eta$ pair correlations as 
$D=4J_{\rm ex}\langle \hat{\boldsymbol{\eta}}_j \cdot \hat{\boldsymbol{\eta}}_\ell\rangle$ 
with $J_{\rm ex}=2\tH^2/U$. Unfortunately, it takes much more computational effort to simulate ${\rm Re}\,\sigma(\omega)$ in the limit $\omega\to 0$ since the larger time simulations with keeping the appropriate accuracy are necessary by iTEBD. This task remains a subject for future work, e.g., by utilizing the exact diagonalization technique~\cite{JChemPhys86}.
See also Ref.~\cite{SM} for the results of the imaginary part of the nonequilibrium optical conductivity, which examine the $1/\omega$ scaling behavior.

\begin{figure}[tb]
  \includegraphics[width=0.99\linewidth]{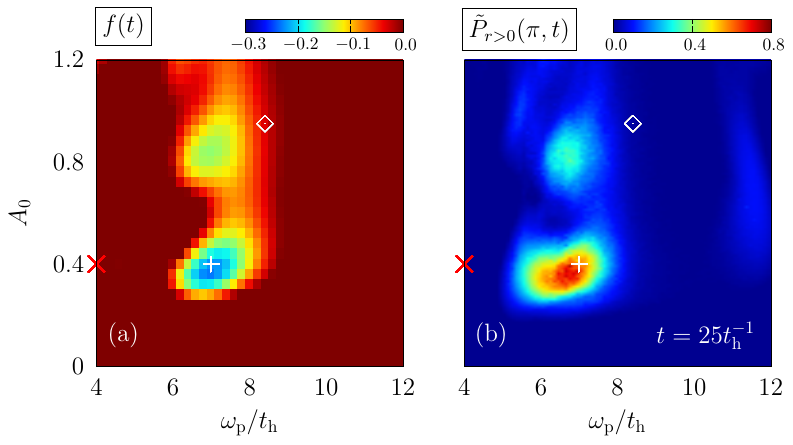}
  \caption{Contour plot of the integrated negative spectral weight $f(t)$ (a) and the modified structure factor $\tilde{P}_{r>0}(q=\pi,t)$ (b) at $t=25\tH^{-1}$ in the $A_0$-$\omega_{\rm p}$ plane for an infinite Hubbard chain at half filling with $U/\tH=8$.
  }
 \label{contour-f}
\end{figure}

Now, we simulate the nonequilibrium optical conductivity $\sigma(\omega; t)$ for various pump-pulse parameters $A_0$ and $\omega_{\rm p}$ and integrate out the spectral weight of $\sigma(\omega; t)$ for $\omega>\Delta_{\rm c}$ as 
$f(t) \equiv \int_{\omega>\Delta_{\rm c}}d\omega\ {\rm Re}\,\sigma(\omega; t)$. 
Figure~\ref{contour-f} demonstrates the contour plot of $f(t)$ of the model~\eqref{hubbard} after pulse irradiation, in which the two-peak structure appears. The position of the highest peak coincides with that of the $\eta$-pair correlations in Fig.~\ref{PpitNd}(a) (marked as `+'), implying the strong relation between the negative conductivity and the formation of $\eta$ pairs due to the population inversion by pump-pulse irradiation. 
Examining the negative conductivity also sharpens the second peak around $A_0=0.84$ and $\omega_{\rm p}/\tH=7.0$, which was unclear in Fig.~\ref{PpitNd}(a). This extra peak can be reconfirmed by exploring the modified structure factor 
$\tilde{P}_{r>0}(q,t)=\sum_{r>0}e^{\mathrm{i}qr}P(r,t)$ in order to get rid of the contribution of local pairs (doublons) from $\tilde{P}(q=\pi,t)$ as demonstrated in Ref.~\cite{Ejima20}. 
In Fig.~\ref{contour-f}(b), we show the contour plot of $P_{r>0}(q=\pi,t)\equiv \sum_{r>0}e^{\mathrm{i}qr}P(r,t)$ in the $A_0$-$\omega_{\rm p}$ plane. Compared with Fig.~\ref{PpitNd}(a), the two-peak structure is more clearly visible. In addition, the positions of the two peaks agree very well with those obtained from Fig.~\ref{contour-f}(a), further indicating that the longer-range correlations play a peculiar role in the $\eta$-pairing dominant regime and are directly related to the negative dynamic conductivity. 

\begin{figure}[tb]
   \includegraphics[width=0.99\linewidth]{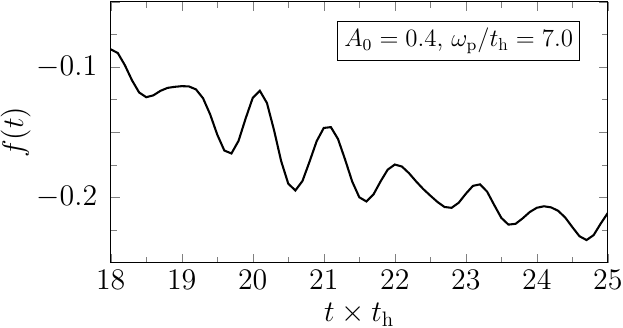}
   \caption{Time dependence of the integrated negative weights $f(t)$ at the $\eta$-pairing dominant point [$A_0=0.4$ and $\omega_{\rm p}/\tH=7.0$ denoted as `+' symbol in Fig.~\ref{contour-f}].
   }
   \label{t-depend-weight}
\end{figure}

Figure~\ref{t-depend-weight} demonstrates the time dependence of the integrated negative weights with the pump-pulse parameters $A_0=0.4$ and $\omega_{\rm p}/\tH=7.0$, i.e., the $\eta$-pairing dominant point denoted as the `+' symbol in Figs.~\ref{PpitNd} and \ref{contour-f}. A clear oscillation can be seen here with the period $T\approx 1/\tH$, which gives us the frequency $\omega^\prime=2\pi/T\simeq6.28$, showing a reasonable agreement with the negative peak position $\omega/\tH\approx6.2$ [see Fig.~\ref{sigma-w}(b)] as well as with the Higgs oscillation in Fig.~\ref{PpitNd}(c). This demonstrates the origin of the negative peak due to the Higgs mode, showing a similar behavior as for the  out-of-equilibrium induced Higgs mode, \cite{annurev:/content/journals/10.1146/annurev-conmatphys-031119-050813,PhysRevLett.13.508, Varma2002,PhysRevLett.109.187002,PhysRevLett.111.057002,Schwarz2020,PhysRevB.101.224510,doi:10.1126/sciadv.1700718}, but with a negative spectral weight, not discussed before. 

The observation of a Higgs-driven negative conductivity peak in the $\eta$-paired regime enables the identification of the $\eta$-pairing-dominated region and provides a direct measure of the Higgs mode frequency via radiation at the Higgs frequency. Similar optical amplification was recently reported for the light-induced superconducting state out of metallic K$_3$C$_{60}$ and requires a prompt quench \cite{PhysRevX.11.011055} which was followed by an observed negative conductivity below $10$ meV but no pronounced peak. In contrast, the $\eta$-paired superconducting state emerges from a Mott insulator, which leads to a negative conductivity peak above the Mott gap. Since the reported radiative Higgs mode is so far exclusive to the 1D case, and does not appear at infinite dimensions~\cite{PhysRevB.102.165136,PhysRevB.108.174515}, it is highly desirable to investigate the optical conductivity after pulse irradiation in between, i.e., in two and three spatial dimensions.

\textit{Acknowledgments} ---
The authors thank K. Sugimoto for fruitful discussions.
This project was made possible by the DLR Quantum Computing Initiative and the Federal Ministry for Economic Aﬀairs and Climate Action; qci.dlr.de/projects/ALQU.
The authors gratefully acknowledge the scientific support and HPC resources provided by the German Aerospace Center (DLR). The HPC system CARO is partially funded by "Ministry of Science and Culture of Lower Saxony" and "Federal Ministry for Economic Affairs and Climate Action".

The iTEBD simulations were performed using the ITensor library~\cite{ITensor}.


\begin{thebibliography}{48}%
\makeatletter
\providecommand \@ifxundefined [1]{%
 \@ifx{#1\undefined}
}%
\providecommand \@ifnum [1]{%
 \ifnum #1\expandafter \@firstoftwo
 \else \expandafter \@secondoftwo
 \fi
}%
\providecommand \@ifx [1]{%
 \ifx #1\expandafter \@firstoftwo
 \else \expandafter \@secondoftwo
 \fi
}%
\providecommand \natexlab [1]{#1}%
\providecommand \enquote  [1]{``#1''}%
\providecommand \bibnamefont  [1]{#1}%
\providecommand \bibfnamefont [1]{#1}%
\providecommand \citenamefont [1]{#1}%
\providecommand \href@noop [0]{\@secondoftwo}%
\providecommand \href [0]{\begingroup \@sanitize@url \@href}%
\providecommand \@href[1]{\@@startlink{#1}\@@href}%
\providecommand \@@href[1]{\endgroup#1\@@endlink}%
\providecommand \@sanitize@url [0]{\catcode `\\12\catcode `\$12\catcode
  `\&12\catcode `\#12\catcode `\^12\catcode `\_12\catcode `\%12\relax}%
\providecommand \@@startlink[1]{}%
\providecommand \@@endlink[0]{}%
\providecommand \url  [0]{\begingroup\@sanitize@url \@url }%
\providecommand \@url [1]{\endgroup\@href {#1}{\urlprefix }}%
\providecommand \urlprefix  [0]{URL }%
\providecommand \Eprint [0]{\href }%
\providecommand \doibase [0]{http://dx.doi.org/}%
\providecommand \selectlanguage [0]{\@gobble}%
\providecommand \bibinfo  [0]{\@secondoftwo}%
\providecommand \bibfield  [0]{\@secondoftwo}%
\providecommand \translation [1]{[#1]}%
\providecommand \BibitemOpen [0]{}%
\providecommand \bibitemStop [0]{}%
\providecommand \bibitemNoStop [0]{.\EOS\space}%
\providecommand \EOS [0]{\spacefactor3000\relax}%
\providecommand \BibitemShut  [1]{\csname bibitem#1\endcsname}%
\let\auto@bib@innerbib\@empty
\bibitem [{\citenamefont {Giannetti}\ \emph {et~al.}(2016)\citenamefont
  {Giannetti}, \citenamefont {Capone}, \citenamefont {Fausti}, \citenamefont
  {Fabrizio}, \citenamefont {Parmigiani},\ and\ \citenamefont
  {Mihailovic}}]{Giannetti2016}%
  \BibitemOpen
  \bibfield  {author} {\bibinfo {author} {\bibfnamefont {C.}~\bibnamefont
  {Giannetti}}, \bibinfo {author} {\bibfnamefont {M.}~\bibnamefont {Capone}},
  \bibinfo {author} {\bibfnamefont {D.}~\bibnamefont {Fausti}}, \bibinfo
  {author} {\bibfnamefont {M.}~\bibnamefont {Fabrizio}}, \bibinfo {author}
  {\bibfnamefont {F.}~\bibnamefont {Parmigiani}}, \ and\ \bibinfo {author}
  {\bibfnamefont {D.}~\bibnamefont {Mihailovic}},\ }\bibinfo {title} {Ultrafast
  optical spectroscopy of strongly correlated materials and high-temperature
  superconductors: a non-equilibrium approach},\ \href {\doibase
  10.1080/00018732.2016.1194044} {\bibfield  {journal} {\bibinfo  {journal}
  {Adv. Phys.}\ }\textbf {\bibinfo {volume} {65}},\ \bibinfo {pages} {58}
  (\bibinfo {year} {2016})}\BibitemShut {NoStop}%
\bibitem [{\citenamefont {Ishihara}(2019)}]{Ishihara2019}%
  \BibitemOpen
  \bibfield  {author} {\bibinfo {author} {\bibfnamefont {S.}~\bibnamefont
  {Ishihara}},\ }\bibinfo {title} {Photoinduced ultrafast phenomena in
  correlated electron magnets},\ \href {\doibase 10.7566/JPSJ.88.072001}
  {\bibfield  {journal} {\bibinfo  {journal} {J. Phys. Soc. Jpn.}\ }\textbf
  {\bibinfo {volume} {88}},\ \bibinfo {pages} {072001} (\bibinfo {year}
  {2019})}\BibitemShut {NoStop}%
\bibitem [{\citenamefont {de~la Torre}\ \emph {et~al.}(2021)\citenamefont
  {de~la Torre}, \citenamefont {Kennes}, \citenamefont {Claassen},
  \citenamefont {Gerber}, \citenamefont {McIver},\ and\ \citenamefont
  {Sentef}}]{RevModPhys.93.041002}%
  \BibitemOpen
  \bibfield  {author} {\bibinfo {author} {\bibfnamefont {A.}~\bibnamefont
  {de~la Torre}}, \bibinfo {author} {\bibfnamefont {D.~M.}\ \bibnamefont
  {Kennes}}, \bibinfo {author} {\bibfnamefont {M.}~\bibnamefont {Claassen}},
  \bibinfo {author} {\bibfnamefont {S.}~\bibnamefont {Gerber}}, \bibinfo
  {author} {\bibfnamefont {J.~W.}\ \bibnamefont {McIver}}, \ and\ \bibinfo
  {author} {\bibfnamefont {M.~A.}\ \bibnamefont {Sentef}},\ }\bibinfo {title}
  {Colloquium: Nonthermal pathways to ultrafast control in quantum materials},\
  \href {\doibase 10.1103/RevModPhys.93.041002} {\bibfield  {journal} {\bibinfo
   {journal} {Rev. Mod. Phys.}\ }\textbf {\bibinfo {volume} {93}},\ \bibinfo
  {pages} {041002} (\bibinfo {year} {2021})}\BibitemShut {NoStop}%
\bibitem [{\citenamefont {Sie}\ \emph {et~al.}(2019)\citenamefont {Sie},
  \citenamefont {Nyby}, \citenamefont {Pemmaraju}, \citenamefont {Park},
  \citenamefont {Shen}, \citenamefont {Yang}, \citenamefont {Hoffmann},
  \citenamefont {Ofori-Okai}, \citenamefont {Li}, \citenamefont {Reid},
  \citenamefont {Weathersby}, \citenamefont {Mannebach}, \citenamefont
  {Finney}, \citenamefont {Rhodes}, \citenamefont {Chenet}, \citenamefont
  {Antony}, \citenamefont {Balicas}, \citenamefont {Hone}, \citenamefont
  {Devereaux}, \citenamefont {Heinz}, \citenamefont {Wang},\ and\ \citenamefont
  {Lindenberg}}]{Sie2019}%
  \BibitemOpen
  \bibfield  {author} {\bibinfo {author} {\bibfnamefont {E.~J.}\ \bibnamefont
  {Sie}}, \bibinfo {author} {\bibfnamefont {C.~M.}\ \bibnamefont {Nyby}},
  \bibinfo {author} {\bibfnamefont {C.~D.}\ \bibnamefont {Pemmaraju}}, \bibinfo
  {author} {\bibfnamefont {S.~J.}\ \bibnamefont {Park}}, \bibinfo {author}
  {\bibfnamefont {X.}~\bibnamefont {Shen}}, \bibinfo {author} {\bibfnamefont
  {J.}~\bibnamefont {Yang}}, \bibinfo {author} {\bibfnamefont {M.~C.}\
  \bibnamefont {Hoffmann}}, \bibinfo {author} {\bibfnamefont {B.~K.}\
  \bibnamefont {Ofori-Okai}}, \bibinfo {author} {\bibfnamefont
  {R.}~\bibnamefont {Li}}, \bibinfo {author} {\bibfnamefont {A.~H.}\
  \bibnamefont {Reid}}, \bibinfo {author} {\bibfnamefont {S.}~\bibnamefont
  {Weathersby}}, \bibinfo {author} {\bibfnamefont {E.}~\bibnamefont
  {Mannebach}}, \bibinfo {author} {\bibfnamefont {N.}~\bibnamefont {Finney}},
  \bibinfo {author} {\bibfnamefont {D.}~\bibnamefont {Rhodes}}, \bibinfo
  {author} {\bibfnamefont {D.}~\bibnamefont {Chenet}}, \bibinfo {author}
  {\bibfnamefont {A.}~\bibnamefont {Antony}}, \bibinfo {author} {\bibfnamefont
  {L.}~\bibnamefont {Balicas}}, \bibinfo {author} {\bibfnamefont
  {J.}~\bibnamefont {Hone}}, \bibinfo {author} {\bibfnamefont {T.~P.}\
  \bibnamefont {Devereaux}}, \bibinfo {author} {\bibfnamefont {T.~F.}\
  \bibnamefont {Heinz}}, \bibinfo {author} {\bibfnamefont {X.}~\bibnamefont
  {Wang}}, \ and\ \bibinfo {author} {\bibfnamefont {A.~M.}\ \bibnamefont
  {Lindenberg}},\ }\bibinfo {title} {An ultrafast symmetry switch in a weyl
  semimetal},\ \href {\doibase 10.1038/s41586-018-0809-4} {\bibfield  {journal}
  {\bibinfo  {journal} {Nature}\ }\textbf {\bibinfo {volume} {565}},\ \bibinfo
  {pages} {61} (\bibinfo {year} {2019})}\BibitemShut {NoStop}%
\bibitem [{\citenamefont {McIver}\ \emph {et~al.}(2019)\citenamefont {McIver},
  \citenamefont {Schulte}, \citenamefont {Stein}, \citenamefont {Matsuyama},
  \citenamefont {Jotzu}, \citenamefont {Meier},\ and\ \citenamefont
  {Cavalleri}}]{McIver2019}%
  \BibitemOpen
  \bibfield  {author} {\bibinfo {author} {\bibfnamefont {J.~W.}\ \bibnamefont
  {McIver}}, \bibinfo {author} {\bibfnamefont {B.}~\bibnamefont {Schulte}},
  \bibinfo {author} {\bibfnamefont {F.-U.}\ \bibnamefont {Stein}}, \bibinfo
  {author} {\bibfnamefont {T.}~\bibnamefont {Matsuyama}}, \bibinfo {author}
  {\bibfnamefont {G.}~\bibnamefont {Jotzu}}, \bibinfo {author} {\bibfnamefont
  {G.}~\bibnamefont {Meier}}, \ and\ \bibinfo {author} {\bibfnamefont
  {A.}~\bibnamefont {Cavalleri}},\ }\bibinfo {title} {Light-induced anomalous
  hall effect in graphene},\ \href {\doibase 10.1038/s41567-019-0698-y}
  {\bibfield  {journal} {\bibinfo  {journal} {Nature Physics}\ }\textbf
  {\bibinfo {volume} {16}},\ \bibinfo {pages} {38–41} (\bibinfo {year}
  {2019})}\BibitemShut {NoStop}%
\bibitem [{\citenamefont {Kogar}\ \emph {et~al.}(2019)\citenamefont {Kogar},
  \citenamefont {Zong}, \citenamefont {Dolgirev}, \citenamefont {Shen},
  \citenamefont {Straquadine}, \citenamefont {Bie}, \citenamefont {Wang},
  \citenamefont {Rohwer}, \citenamefont {Tung}, \citenamefont {Yang},
  \citenamefont {Li}, \citenamefont {Yang}, \citenamefont {Weathersby},
  \citenamefont {Park}, \citenamefont {Kozina}, \citenamefont {Sie},
  \citenamefont {Wen}, \citenamefont {Jarillo-Herrero}, \citenamefont {Fisher},
  \citenamefont {Wang},\ and\ \citenamefont {Gedik}}]{Kogar2019}%
  \BibitemOpen
  \bibfield  {author} {\bibinfo {author} {\bibfnamefont {A.}~\bibnamefont
  {Kogar}}, \bibinfo {author} {\bibfnamefont {A.}~\bibnamefont {Zong}},
  \bibinfo {author} {\bibfnamefont {P.~E.}\ \bibnamefont {Dolgirev}}, \bibinfo
  {author} {\bibfnamefont {X.}~\bibnamefont {Shen}}, \bibinfo {author}
  {\bibfnamefont {J.}~\bibnamefont {Straquadine}}, \bibinfo {author}
  {\bibfnamefont {Y.-Q.}\ \bibnamefont {Bie}}, \bibinfo {author} {\bibfnamefont
  {X.}~\bibnamefont {Wang}}, \bibinfo {author} {\bibfnamefont {T.}~\bibnamefont
  {Rohwer}}, \bibinfo {author} {\bibfnamefont {I.-C.}\ \bibnamefont {Tung}},
  \bibinfo {author} {\bibfnamefont {Y.}~\bibnamefont {Yang}}, \bibinfo {author}
  {\bibfnamefont {R.}~\bibnamefont {Li}}, \bibinfo {author} {\bibfnamefont
  {J.}~\bibnamefont {Yang}}, \bibinfo {author} {\bibfnamefont {S.}~\bibnamefont
  {Weathersby}}, \bibinfo {author} {\bibfnamefont {S.}~\bibnamefont {Park}},
  \bibinfo {author} {\bibfnamefont {M.~E.}\ \bibnamefont {Kozina}}, \bibinfo
  {author} {\bibfnamefont {E.~J.}\ \bibnamefont {Sie}}, \bibinfo {author}
  {\bibfnamefont {H.}~\bibnamefont {Wen}}, \bibinfo {author} {\bibfnamefont
  {P.}~\bibnamefont {Jarillo-Herrero}}, \bibinfo {author} {\bibfnamefont
  {I.~R.}\ \bibnamefont {Fisher}}, \bibinfo {author} {\bibfnamefont
  {X.}~\bibnamefont {Wang}}, \ and\ \bibinfo {author} {\bibfnamefont
  {N.}~\bibnamefont {Gedik}},\ }\bibinfo {title} {Light-induced charge density
  wave in LaTe$_3$},\ \href {\doibase 10.1038/s41567-019-0705-3} {\bibfield
  {journal} {\bibinfo  {journal} {Nat. Phys.}\ }\textbf {\bibinfo {volume}
  {16}},\ \bibinfo {pages} {159} (\bibinfo {year} {2019})}\BibitemShut
  {NoStop}%
\bibitem [{\citenamefont {Basov}\ \emph {et~al.}(2017)\citenamefont {Basov},
  \citenamefont {Averitt},\ and\ \citenamefont {Hsieh}}]{Basov2017}%
  \BibitemOpen
  \bibfield  {author} {\bibinfo {author} {\bibfnamefont {D.~N.}\ \bibnamefont
  {Basov}}, \bibinfo {author} {\bibfnamefont {R.~D.}\ \bibnamefont {Averitt}},
  \ and\ \bibinfo {author} {\bibfnamefont {D.}~\bibnamefont {Hsieh}},\
  }\bibinfo {title} {Towards properties on demand in quantum materials},\ \href
  {\doibase 10.1038/nmat5017} {\bibfield  {journal} {\bibinfo  {journal} {Nat.
  Mater.}\ }\textbf {\bibinfo {volume} {16}},\ \bibinfo {pages} {1077}
  (\bibinfo {year} {2017})}\BibitemShut {NoStop}%
\bibitem [{\citenamefont {Mankowsky}\ \emph {et~al.}(2014)\citenamefont
  {Mankowsky}, \citenamefont {Subedi}, \citenamefont {F{\"o}rst}, \citenamefont
  {Mariager}, \citenamefont {Chollet}, \citenamefont {Lemke}, \citenamefont
  {Robinson}, \citenamefont {Glownia}, \citenamefont {Minitti}, \citenamefont
  {Frano}, \citenamefont {Fechner}, \citenamefont {Spaldin}, \citenamefont
  {Loew}, \citenamefont {Keimer}, \citenamefont {Georges},\ and\ \citenamefont
  {Cavalleri}}]{Mankowsky2014}%
  \BibitemOpen
  \bibfield  {author} {\bibinfo {author} {\bibfnamefont {R.}~\bibnamefont
  {Mankowsky}}, \bibinfo {author} {\bibfnamefont {A.}~\bibnamefont {Subedi}},
  \bibinfo {author} {\bibfnamefont {M.}~\bibnamefont {F{\"o}rst}}, \bibinfo
  {author} {\bibfnamefont {S.~O.}\ \bibnamefont {Mariager}}, \bibinfo {author}
  {\bibfnamefont {M.}~\bibnamefont {Chollet}}, \bibinfo {author} {\bibfnamefont
  {H.~T.}\ \bibnamefont {Lemke}}, \bibinfo {author} {\bibfnamefont {J.~S.}\
  \bibnamefont {Robinson}}, \bibinfo {author} {\bibfnamefont {J.~M.}\
  \bibnamefont {Glownia}}, \bibinfo {author} {\bibfnamefont {M.~P.}\
  \bibnamefont {Minitti}}, \bibinfo {author} {\bibfnamefont {A.}~\bibnamefont
  {Frano}}, \bibinfo {author} {\bibfnamefont {M.}~\bibnamefont {Fechner}},
  \bibinfo {author} {\bibfnamefont {N.~A.}\ \bibnamefont {Spaldin}}, \bibinfo
  {author} {\bibfnamefont {T.}~\bibnamefont {Loew}}, \bibinfo {author}
  {\bibfnamefont {B.}~\bibnamefont {Keimer}}, \bibinfo {author} {\bibfnamefont
  {A.}~\bibnamefont {Georges}}, \ and\ \bibinfo {author} {\bibfnamefont
  {A.}~\bibnamefont {Cavalleri}},\ }\bibinfo {title} {Nonlinear lattice
  dynamics as a basis for enhanced superconductivity in YBa$_2$Cu$_3$O$_{6.5}$},\ \href
  {\doibase 10.1038/nature13875} {\bibfield  {journal} {\bibinfo  {journal}
  {Nature}\ }\textbf {\bibinfo {volume} {516}},\ \bibinfo {pages} {71}
  (\bibinfo {year} {2014})}\BibitemShut {NoStop}%
\bibitem [{\citenamefont {Mitrano}\ \emph {et~al.}(2016)\citenamefont
  {Mitrano}, \citenamefont {Cantaluppi}, \citenamefont {Nicoletti},
  \citenamefont {Kaiser}, \citenamefont {Perucchi}, \citenamefont {Lupi},
  \citenamefont {Di~Pietro}, \citenamefont {Pontiroli}, \citenamefont {Riccò},
  \citenamefont {Clark}, \citenamefont {Jaksch},\ and\ \citenamefont
  {Cavalleri}}]{Mitrano2016}%
  \BibitemOpen
  \bibfield  {author} {\bibinfo {author} {\bibfnamefont {M.}~\bibnamefont
  {Mitrano}}, \bibinfo {author} {\bibfnamefont {A.}~\bibnamefont {Cantaluppi}},
  \bibinfo {author} {\bibfnamefont {D.}~\bibnamefont {Nicoletti}}, \bibinfo
  {author} {\bibfnamefont {S.}~\bibnamefont {Kaiser}}, \bibinfo {author}
  {\bibfnamefont {A.}~\bibnamefont {Perucchi}}, \bibinfo {author}
  {\bibfnamefont {S.}~\bibnamefont {Lupi}}, \bibinfo {author} {\bibfnamefont
  {P.}~\bibnamefont {Di~Pietro}}, \bibinfo {author} {\bibfnamefont
  {D.}~\bibnamefont {Pontiroli}}, \bibinfo {author} {\bibfnamefont
  {M.}~\bibnamefont {Riccò}}, \bibinfo {author} {\bibfnamefont {S.~R.}\
  \bibnamefont {Clark}}, \bibinfo {author} {\bibfnamefont {D.}~\bibnamefont
  {Jaksch}}, \ and\ \bibinfo {author} {\bibfnamefont {A.}~\bibnamefont
  {Cavalleri}},\ }\bibinfo {title} {Possible light-induced superconductivity in
  K$_3$C$_{60}$ at high temperature},\ \href {\doibase 10.1038/nature16522} {\bibfield
  {journal} {\bibinfo  {journal} {Nature}\ }\textbf {\bibinfo {volume} {530}},\
  \bibinfo {pages} {461–464} (\bibinfo {year} {2016})}\BibitemShut {NoStop}%
\bibitem [{\citenamefont {Buzzi}\ \emph {et~al.}(2020)\citenamefont {Buzzi},
  \citenamefont {Nicoletti}, \citenamefont {Fechner}, \citenamefont
  {Tancogne-Dejean}, \citenamefont {Sentef}, \citenamefont {Georges},
  \citenamefont {Biesner}, \citenamefont {Uykur}, \citenamefont {Dressel},
  \citenamefont {Henderson}, \citenamefont {Siegrist}, \citenamefont
  {Schlueter}, \citenamefont {Miyagawa}, \citenamefont {Kanoda}, \citenamefont
  {Nam}, \citenamefont {Ardavan}, \citenamefont {Coulthard}, \citenamefont
  {Tindall}, \citenamefont {Schlawin}, \citenamefont {Jaksch},\ and\
  \citenamefont {Cavalleri}}]{PhysRevX.10.031028}%
  \BibitemOpen
  \bibfield  {author} {\bibinfo {author} {\bibfnamefont {M.}~\bibnamefont
  {Buzzi}}, \bibinfo {author} {\bibfnamefont {D.}~\bibnamefont {Nicoletti}},
  \bibinfo {author} {\bibfnamefont {M.}~\bibnamefont {Fechner}}, \bibinfo
  {author} {\bibfnamefont {N.}~\bibnamefont {Tancogne-Dejean}}, \bibinfo
  {author} {\bibfnamefont {M.~A.}\ \bibnamefont {Sentef}}, \bibinfo {author}
  {\bibfnamefont {A.}~\bibnamefont {Georges}}, \bibinfo {author} {\bibfnamefont
  {T.}~\bibnamefont {Biesner}}, \bibinfo {author} {\bibfnamefont
  {E.}~\bibnamefont {Uykur}}, \bibinfo {author} {\bibfnamefont
  {M.}~\bibnamefont {Dressel}}, \bibinfo {author} {\bibfnamefont
  {A.}~\bibnamefont {Henderson}}, \bibinfo {author} {\bibfnamefont
  {T.}~\bibnamefont {Siegrist}}, \bibinfo {author} {\bibfnamefont {J.~A.}\
  \bibnamefont {Schlueter}}, \bibinfo {author} {\bibfnamefont {K.}~\bibnamefont
  {Miyagawa}}, \bibinfo {author} {\bibfnamefont {K.}~\bibnamefont {Kanoda}},
  \bibinfo {author} {\bibfnamefont {M.-S.}\ \bibnamefont {Nam}}, \bibinfo
  {author} {\bibfnamefont {A.}~\bibnamefont {Ardavan}}, \bibinfo {author}
  {\bibfnamefont {J.}~\bibnamefont {Coulthard}}, \bibinfo {author}
  {\bibfnamefont {J.}~\bibnamefont {Tindall}}, \bibinfo {author} {\bibfnamefont
  {F.}~\bibnamefont {Schlawin}}, \bibinfo {author} {\bibfnamefont
  {D.}~\bibnamefont {Jaksch}}, \ and\ \bibinfo {author} {\bibfnamefont
  {A.}~\bibnamefont {Cavalleri}},\ }\bibinfo {title} {Photomolecular
  high-temperature superconductivity},\ \href {\doibase
  10.1103/PhysRevX.10.031028} {\bibfield  {journal} {\bibinfo  {journal} {Phys.
  Rev. X}\ }\textbf {\bibinfo {volume} {10}},\ \bibinfo {pages} {031028}
  (\bibinfo {year} {2020})}\BibitemShut {NoStop}%
\bibitem [{\citenamefont {Budden}\ \emph {et~al.}(2021)\citenamefont {Budden},
  \citenamefont {Gebert}, \citenamefont {Buzzi}, \citenamefont {Jotzu},
  \citenamefont {Wang}, \citenamefont {Matsuyama}, \citenamefont {Meier},
  \citenamefont {Laplace}, \citenamefont {Pontiroli}, \citenamefont {Riccò},
  \citenamefont {Schlawin}, \citenamefont {Jaksch},\ and\ \citenamefont
  {Cavalleri}}]{Budden2021}%
  \BibitemOpen
  \bibfield  {author} {\bibinfo {author} {\bibfnamefont {M.}~\bibnamefont
  {Budden}}, \bibinfo {author} {\bibfnamefont {T.}~\bibnamefont {Gebert}},
  \bibinfo {author} {\bibfnamefont {M.}~\bibnamefont {Buzzi}}, \bibinfo
  {author} {\bibfnamefont {G.}~\bibnamefont {Jotzu}}, \bibinfo {author}
  {\bibfnamefont {E.}~\bibnamefont {Wang}}, \bibinfo {author} {\bibfnamefont
  {T.}~\bibnamefont {Matsuyama}}, \bibinfo {author} {\bibfnamefont
  {G.}~\bibnamefont {Meier}}, \bibinfo {author} {\bibfnamefont
  {Y.}~\bibnamefont {Laplace}}, \bibinfo {author} {\bibfnamefont
  {D.}~\bibnamefont {Pontiroli}}, \bibinfo {author} {\bibfnamefont
  {M.}~\bibnamefont {Riccò}}, \bibinfo {author} {\bibfnamefont
  {F.}~\bibnamefont {Schlawin}}, \bibinfo {author} {\bibfnamefont
  {D.}~\bibnamefont {Jaksch}}, \ and\ \bibinfo {author} {\bibfnamefont
  {A.}~\bibnamefont {Cavalleri}},\ }\bibinfo {title} {Evidence for metastable
  photo-induced superconductivity in K$_3$C$_{60}$},\ \href {\doibase
  10.1038/s41567-020-01148-1} {\bibfield  {journal} {\bibinfo  {journal} {Nat.
  Phys.}\ }\textbf {\bibinfo {volume} {17}},\ \bibinfo {pages} {611} (\bibinfo
  {year} {2021})}\BibitemShut {NoStop}%
\bibitem [{\citenamefont {Eckstein}\ \emph {et~al.}(2010)\citenamefont
  {Eckstein}, \citenamefont {Kollar},\ and\ \citenamefont
  {Werner}}]{PhysRevB.81.115131}%
  \BibitemOpen
  \bibfield  {author} {\bibinfo {author} {\bibfnamefont {M.}~\bibnamefont
  {Eckstein}}, \bibinfo {author} {\bibfnamefont {M.}~\bibnamefont {Kollar}}, \
  and\ \bibinfo {author} {\bibfnamefont {P.}~\bibnamefont {Werner}},\ }\bibinfo
  {title} {Interaction quench in the Hubbard model: Relaxation of the spectral
  function and the optical conductivity},\ \href {\doibase
  10.1103/PhysRevB.81.115131} {\bibfield  {journal} {\bibinfo  {journal} {Phys.
  Rev. B}\ }\textbf {\bibinfo {volume} {81}},\ \bibinfo {pages} {115131}
  (\bibinfo {year} {2010})}\BibitemShut {NoStop}%
\bibitem [{\citenamefont {Bittner}\ \emph {et~al.}(2019)\citenamefont
  {Bittner}, \citenamefont {Tohyama}, \citenamefont {Kaiser},\ and\
  \citenamefont {Manske}}]{doi:10.7566/JPSJ.88.044704}%
  \BibitemOpen
  \bibfield  {author} {\bibinfo {author} {\bibfnamefont {N.}~\bibnamefont
  {Bittner}}, \bibinfo {author} {\bibfnamefont {T.}~\bibnamefont {Tohyama}},
  \bibinfo {author} {\bibfnamefont {S.}~\bibnamefont {Kaiser}}, \ and\ \bibinfo
  {author} {\bibfnamefont {D.}~\bibnamefont {Manske}},\ }\bibinfo {title}
  {Possible light-induced superconductivity in a strongly correlated electron
  system},\ \href {\doibase 10.7566/JPSJ.88.044704} {\bibfield  {journal}
  {\bibinfo  {journal} {J. Phys. Soc. Jpn.}\ }\textbf {\bibinfo {volume}
  {88}},\ \bibinfo {pages} {044704} (\bibinfo {year} {2019})}\BibitemShut
  {NoStop}%
\bibitem [{\citenamefont {Kennes}\ \emph {et~al.}(2017)\citenamefont {Kennes},
  \citenamefont {Wilner}, \citenamefont {Reichman},\ and\ \citenamefont
  {Millis}}]{PhysRevB.96.054506}%
  \BibitemOpen
  \bibfield  {author} {\bibinfo {author} {\bibfnamefont {D.~M.}\ \bibnamefont
  {Kennes}}, \bibinfo {author} {\bibfnamefont {E.~Y.}\ \bibnamefont {Wilner}},
  \bibinfo {author} {\bibfnamefont {D.~R.}\ \bibnamefont {Reichman}}, \ and\
  \bibinfo {author} {\bibfnamefont {A.~J.}\ \bibnamefont {Millis}},\ }\bibinfo
  {title} {Nonequilibrium optical conductivity: General theory and application
  to transient phases},\ \href {\doibase 10.1103/PhysRevB.96.054506} {\bibfield
   {journal} {\bibinfo  {journal} {Phys. Rev. B}\ }\textbf {\bibinfo {volume}
  {96}},\ \bibinfo {pages} {054506} (\bibinfo {year} {2017})}\BibitemShut
  {NoStop}%
\bibitem [{\citenamefont {Paeckel}\ \emph {et~al.}(2020)\citenamefont
  {Paeckel}, \citenamefont {Fauseweh}, \citenamefont {Osterkorn}, \citenamefont
  {K\"ohler}, \citenamefont {Manske},\ and\ \citenamefont
  {Manmana}}]{PhysRevB.101.180507}%
  \BibitemOpen
  \bibfield  {author} {\bibinfo {author} {\bibfnamefont {S.}~\bibnamefont
  {Paeckel}}, \bibinfo {author} {\bibfnamefont {B.}~\bibnamefont {Fauseweh}},
  \bibinfo {author} {\bibfnamefont {A.}~\bibnamefont {Osterkorn}}, \bibinfo
  {author} {\bibfnamefont {T.}~\bibnamefont {K\"ohler}}, \bibinfo {author}
  {\bibfnamefont {D.}~\bibnamefont {Manske}}, \ and\ \bibinfo {author}
  {\bibfnamefont {S.~R.}\ \bibnamefont {Manmana}},\ }\bibinfo {title}
  {Detecting superconductivity out of equilibrium},\ \href {\doibase
  10.1103/PhysRevB.101.180507} {\bibfield  {journal} {\bibinfo  {journal}
  {Phys. Rev. B}\ }\textbf {\bibinfo {volume} {101}},\ \bibinfo {pages}
  {180507} (\bibinfo {year} {2020})}\BibitemShut {NoStop}%
\bibitem [{\citenamefont {Marten}\ \emph {et~al.}(2023)\citenamefont {Marten},
  \citenamefont {Bollmark}, \citenamefont {Köhler}, \citenamefont {Manmana},\
  and\ \citenamefont {Kantian}}]{10.21468/SciPostPhys.15.6.236}%
  \BibitemOpen
  \bibfield  {author} {\bibinfo {author} {\bibfnamefont {S.}~\bibnamefont
  {Marten}}, \bibinfo {author} {\bibfnamefont {G.}~\bibnamefont {Bollmark}},
  \bibinfo {author} {\bibfnamefont {T.}~\bibnamefont {Köhler}}, \bibinfo
  {author} {\bibfnamefont {S.~R.}\ \bibnamefont {Manmana}}, \ and\ \bibinfo
  {author} {\bibfnamefont {A.}~\bibnamefont {Kantian}},\ }\bibinfo {title}
  {{Transient superconductivity in three-dimensional {H}ubbard systems by
  combining matrix-product states and self-consistent mean-field theory}},\
  \href {\doibase 10.21468/SciPostPhys.15.6.236} {\bibfield  {journal}
  {\bibinfo  {journal} {SciPost Phys.}\ }\textbf {\bibinfo {volume} {15}},\
  \bibinfo {pages} {236} (\bibinfo {year} {2023})}\BibitemShut {NoStop}%
\bibitem [{\citenamefont {Yang}(1989)}]{PhysRevLett.63.2144}%
  \BibitemOpen
  \bibfield  {author} {\bibinfo {author} {\bibfnamefont {C.~N.}\ \bibnamefont
  {Yang}},\ }\bibinfo {title} {\ensuremath{\eta} pairing and off-diagonal
  long-range order in a {H}ubbard model},\ \href {\doibase
  10.1103/PhysRevLett.63.2144} {\bibfield  {journal} {\bibinfo  {journal}
  {Phys. Rev. Lett.}\ }\textbf {\bibinfo {volume} {63}},\ \bibinfo {pages}
  {2144} (\bibinfo {year} {1989})}\BibitemShut {NoStop}%
\bibitem [{\citenamefont {Kaneko}\ \emph {et~al.}(2019)\citenamefont {Kaneko},
  \citenamefont {Shirakawa}, \citenamefont {Sorella},\ and\ \citenamefont
  {Yunoki}}]{Kaneko19}%
  \BibitemOpen
  \bibfield  {author} {\bibinfo {author} {\bibfnamefont {T.}~\bibnamefont
  {Kaneko}}, \bibinfo {author} {\bibfnamefont {T.}~\bibnamefont {Shirakawa}},
  \bibinfo {author} {\bibfnamefont {S.}~\bibnamefont {Sorella}}, \ and\
  \bibinfo {author} {\bibfnamefont {S.}~\bibnamefont {Yunoki}},\ }\bibinfo
  {title} {Photoinduced $\ensuremath{\eta}$ pairing in the {H}ubbard model},\
  \href {\doibase 10.1103/PhysRevLett.122.077002} {\bibfield  {journal}
  {\bibinfo  {journal} {Phys. Rev. Lett.}\ }\textbf {\bibinfo {volume} {122}},\
  \bibinfo {pages} {077002} (\bibinfo {year} {2019})}\BibitemShut {NoStop}%
\bibitem [{\citenamefont {Tindall}\ \emph {et~al.}(2020)\citenamefont
  {Tindall}, \citenamefont {Schlawin}, \citenamefont {Buzzi}, \citenamefont
  {Nicoletti}, \citenamefont {Coulthard}, \citenamefont {Gao}, \citenamefont
  {Cavalleri}, \citenamefont {Sentef},\ and\ \citenamefont
  {Jaksch}}]{PhysRevLett.125.137001}%
  \BibitemOpen
  \bibfield  {author} {\bibinfo {author} {\bibfnamefont {J.}~\bibnamefont
  {Tindall}}, \bibinfo {author} {\bibfnamefont {F.}~\bibnamefont {Schlawin}},
  \bibinfo {author} {\bibfnamefont {M.}~\bibnamefont {Buzzi}}, \bibinfo
  {author} {\bibfnamefont {D.}~\bibnamefont {Nicoletti}}, \bibinfo {author}
  {\bibfnamefont {J.~R.}\ \bibnamefont {Coulthard}}, \bibinfo {author}
  {\bibfnamefont {H.}~\bibnamefont {Gao}}, \bibinfo {author} {\bibfnamefont
  {A.}~\bibnamefont {Cavalleri}}, \bibinfo {author} {\bibfnamefont {M.~A.}\
  \bibnamefont {Sentef}}, \ and\ \bibinfo {author} {\bibfnamefont
  {D.}~\bibnamefont {Jaksch}},\ }\bibinfo {title} {Dynamical order and
  superconductivity in a frustrated many-body system},\ \href {\doibase
  10.1103/PhysRevLett.125.137001} {\bibfield  {journal} {\bibinfo  {journal}
  {Phys. Rev. Lett.}\ }\textbf {\bibinfo {volume} {125}},\ \bibinfo {pages}
  {137001} (\bibinfo {year} {2020})}\BibitemShut {NoStop}%
\bibitem [{\citenamefont {Chen}\ \emph {et~al.}(2021)\citenamefont {Chen},
  \citenamefont {Wang}, \citenamefont {Rebec}, \citenamefont {Jia},
  \citenamefont {Hashimoto}, \citenamefont {Lu}, \citenamefont {Moritz},
  \citenamefont {Moore}, \citenamefont {Devereaux},\ and\ \citenamefont
  {Shen}}]{science.373.1235}%
  \BibitemOpen
  \bibfield  {author} {\bibinfo {author} {\bibfnamefont {Z.}~\bibnamefont
  {Chen}}, \bibinfo {author} {\bibfnamefont {Y.}~\bibnamefont {Wang}}, \bibinfo
  {author} {\bibfnamefont {S.~N.}\ \bibnamefont {Rebec}}, \bibinfo {author}
  {\bibfnamefont {T.}~\bibnamefont {Jia}}, \bibinfo {author} {\bibfnamefont
  {M.}~\bibnamefont {Hashimoto}}, \bibinfo {author} {\bibfnamefont
  {D.}~\bibnamefont {Lu}}, \bibinfo {author} {\bibfnamefont {B.}~\bibnamefont
  {Moritz}}, \bibinfo {author} {\bibfnamefont {R.~G.}\ \bibnamefont {Moore}},
  \bibinfo {author} {\bibfnamefont {T.~P.}\ \bibnamefont {Devereaux}}, \ and\
  \bibinfo {author} {\bibfnamefont {Z.-X.}\ \bibnamefont {Shen}},\ }\bibinfo
  {title} {Anomalously strong near-neighbor attraction in doped 1D cuprate
  chains},\ \href {\doibase 10.1126/science.abf5174} {\bibfield  {journal}
  {\bibinfo  {journal} {Science}\ }\textbf {\bibinfo {volume} {373}},\ \bibinfo
  {pages} {1235} (\bibinfo {year} {2021})}\BibitemShut {NoStop}%
\bibitem [{\citenamefont {Matsukawa}\ \emph {et~al.}(2004)\citenamefont
  {Matsukawa}, \citenamefont {Yamada}, \citenamefont {Chiba}, \citenamefont
  {Ogasawara}, \citenamefont {Shibata}, \citenamefont {Matsushita},\ and\
  \citenamefont {Takano}}]{MATSUKAWA2004101}%
  \BibitemOpen
  \bibfield  {author} {\bibinfo {author} {\bibfnamefont {M.}~\bibnamefont
  {Matsukawa}}, \bibinfo {author} {\bibfnamefont {Y.}~\bibnamefont {Yamada}},
  \bibinfo {author} {\bibfnamefont {M.}~\bibnamefont {Chiba}}, \bibinfo
  {author} {\bibfnamefont {H.}~\bibnamefont {Ogasawara}}, \bibinfo {author}
  {\bibfnamefont {T.}~\bibnamefont {Shibata}}, \bibinfo {author} {\bibfnamefont
  {A.}~\bibnamefont {Matsushita}}, \ and\ \bibinfo {author} {\bibfnamefont
  {Y.}~\bibnamefont {Takano}},\ }\bibinfo {title} {Superconductivity in
  Pr$_2$Ba$_4$Cu$_{7}$O$_{15-\delta}$ with metallic double chains},\ \href
  {\doibase https://doi.org/10.1016/j.physc.2004.07.002} {\bibfield  {journal}
  {\bibinfo  {journal} {Physica C}\ }\textbf {\bibinfo {volume} {411}},\
  \bibinfo {pages} {101} (\bibinfo {year} {2004})}\BibitemShut {NoStop}%
\bibitem [{\citenamefont {Habaguchi}\ \emph {et~al.}(2011)\citenamefont
  {Habaguchi}, \citenamefont {Ōno}, \citenamefont {Ying Du~Gh}, \citenamefont
  {Sano},\ and\ \citenamefont {Yamada}}]{JPSJ.80.024708}%
  \BibitemOpen
  \bibfield  {author} {\bibinfo {author} {\bibfnamefont {T.}~\bibnamefont
  {Habaguchi}}, \bibinfo {author} {\bibfnamefont {Y.}~\bibnamefont {Ōno}},
  \bibinfo {author} {\bibfnamefont {H.}~\bibnamefont {Ying Du~Gh}}, \bibinfo
  {author} {\bibfnamefont {K.}~\bibnamefont {Sano}}, \ and\ \bibinfo {author}
  {\bibfnamefont {Y.}~\bibnamefont {Yamada}},\ }\bibinfo {title} {Electronic
  states and superconducting transition temperature based on the
  Tomonaga–Luttinger liquid in {Pr$_2$Ba$_4$Cu$_7$O$_{15-\delta}$}},\ \href
  {\doibase 10.1143/JPSJ.80.024708} {\bibfield  {journal} {\bibinfo  {journal}
  {J. Phys. Soc. Jpn.}\ }\textbf {\bibinfo {volume} {80}},\ \bibinfo {pages}
  {024708} (\bibinfo {year} {2011})}\BibitemShut {NoStop}%
\bibitem [{\citenamefont {Kaneko}\ \emph {et~al.}(2020)\citenamefont {Kaneko},
  \citenamefont {Yunoki},\ and\ \citenamefont {Millis}}]{Kaneko20}%
  \BibitemOpen
  \bibfield  {author} {\bibinfo {author} {\bibfnamefont {T.}~\bibnamefont
  {Kaneko}}, \bibinfo {author} {\bibfnamefont {S.}~\bibnamefont {Yunoki}}, \
  and\ \bibinfo {author} {\bibfnamefont {A.~J.}\ \bibnamefont {Millis}},\
  }\bibinfo {title} {Charge stiffness and long-range correlation in the
  optically induced $\ensuremath{\eta}$-pairing state of the one-dimensional
  Hubbard model},\ \href {\doibase 10.1103/PhysRevResearch.2.032027} {\bibfield
   {journal} {\bibinfo  {journal} {Phys. Rev. Research}\ }\textbf {\bibinfo
  {volume} {2}},\ \bibinfo {pages} {032027} (\bibinfo {year}
  {2020})}\BibitemShut {NoStop}%
\bibitem [{\citenamefont {Ejima}\ \emph {et~al.}(2022)\citenamefont {Ejima},
  \citenamefont {Lange},\ and\ \citenamefont {Fehske}}]{Ejima22}%
  \BibitemOpen
  \bibfield  {author} {\bibinfo {author} {\bibfnamefont {S.}~\bibnamefont
  {Ejima}}, \bibinfo {author} {\bibfnamefont {F.}~\bibnamefont {Lange}}, \ and\
  \bibinfo {author} {\bibfnamefont {H.}~\bibnamefont {Fehske}},\ }\bibinfo
  {title} {Nonequilibrium dynamics in pumped Mott insulators},\ \href {\doibase
  10.1103/PhysRevResearch.4.L012012} {\bibfield  {journal} {\bibinfo  {journal}
  {Phys. Rev. Research}\ }\textbf {\bibinfo {volume} {4}},\ \bibinfo {pages}
  {L012012} (\bibinfo {year} {2022})}\BibitemShut {NoStop}%
\bibitem [{\citenamefont {Ejima}\ \emph {et~al.}(2023)\citenamefont {Ejima},
  \citenamefont {Lange},\ and\ \citenamefont {Fehske}}]{Ejima23}%
  \BibitemOpen
  \bibfield  {author} {\bibinfo {author} {\bibfnamefont {S.}~\bibnamefont
  {Ejima}}, \bibinfo {author} {\bibfnamefont {F.}~\bibnamefont {Lange}}, \ and\
  \bibinfo {author} {\bibfnamefont {H.}~\bibnamefont {Fehske}},\ }\bibinfo
  {title} {Entanglement analysis of photoinduced $\eta$-pairing states},\ \href
  {\doibase 10.1140/epjs/s11734-023-00975-6} {\bibfield  {journal} {\bibinfo
  {journal} {Eur. Phys. J. Spec. Top.}\ }\textbf {\bibinfo {volume} {232}},\
  \bibinfo {pages} {3479} (\bibinfo {year} {2023})}\BibitemShut {NoStop}%
\bibitem [{\citenamefont {Islam}\ \emph {et~al.}(2015)\citenamefont {Islam},
  \citenamefont {Ma}, \citenamefont {Preiss}, \citenamefont {Eric~Tai},
  \citenamefont {Lukin}, \citenamefont {Rispoli},\ and\ \citenamefont
  {Greiner}}]{Islam2015}%
  \BibitemOpen
  \bibfield  {author} {\bibinfo {author} {\bibfnamefont {R.}~\bibnamefont
  {Islam}}, \bibinfo {author} {\bibfnamefont {R.}~\bibnamefont {Ma}}, \bibinfo
  {author} {\bibfnamefont {P.~M.}\ \bibnamefont {Preiss}}, \bibinfo {author}
  {\bibfnamefont {M.}~\bibnamefont {Eric~Tai}}, \bibinfo {author}
  {\bibfnamefont {A.}~\bibnamefont {Lukin}}, \bibinfo {author} {\bibfnamefont
  {M.}~\bibnamefont {Rispoli}}, \ and\ \bibinfo {author} {\bibfnamefont
  {M.}~\bibnamefont {Greiner}},\ }\bibinfo {title} {Measuring entanglement
  entropy in a quantum many-body system},\ \href {\doibase 10.1038/nature15750}
  {\bibfield  {journal} {\bibinfo  {journal} {Nature}\ }\textbf {\bibinfo
  {volume} {528}},\ \bibinfo {pages} {77} (\bibinfo {year} {2015})}\BibitemShut
  {NoStop}%
\bibitem [{\citenamefont {Kaufman}\ \emph {et~al.}(2016)\citenamefont
  {Kaufman}, \citenamefont {Tai}, \citenamefont {Lukin}, \citenamefont
  {Rispoli}, \citenamefont {Schittko}, \citenamefont {Preiss},\ and\
  \citenamefont {Greiner}}]{Kaufman2016}%
  \BibitemOpen
  \bibfield  {author} {\bibinfo {author} {\bibfnamefont {A.~M.}\ \bibnamefont
  {Kaufman}}, \bibinfo {author} {\bibfnamefont {M.~E.}\ \bibnamefont {Tai}},
  \bibinfo {author} {\bibfnamefont {A.}~\bibnamefont {Lukin}}, \bibinfo
  {author} {\bibfnamefont {M.}~\bibnamefont {Rispoli}}, \bibinfo {author}
  {\bibfnamefont {R.}~\bibnamefont {Schittko}}, \bibinfo {author}
  {\bibfnamefont {P.~M.}\ \bibnamefont {Preiss}}, \ and\ \bibinfo {author}
  {\bibfnamefont {M.}~\bibnamefont {Greiner}},\ }\bibinfo {title} {Quantum
  thermalization through entanglement in an isolated many-body system},\ \href
  {\doibase 10.1126/science.aaf6725} {\bibfield  {journal} {\bibinfo  {journal}
  {Science}\ }\textbf {\bibinfo {volume} {353}},\ \bibinfo {pages} {794}
  (\bibinfo {year} {2016})}\BibitemShut {NoStop}%
\bibitem [{\citenamefont {Rosch}\ \emph {et~al.}(2008)\citenamefont {Rosch},
  \citenamefont {Rasch}, \citenamefont {Binz},\ and\ \citenamefont
  {Vojta}}]{PhysRevLett.101.265301}%
  \BibitemOpen
  \bibfield  {author} {\bibinfo {author} {\bibfnamefont {A.}~\bibnamefont
  {Rosch}}, \bibinfo {author} {\bibfnamefont {D.}~\bibnamefont {Rasch}},
  \bibinfo {author} {\bibfnamefont {B.}~\bibnamefont {Binz}}, \ and\ \bibinfo
  {author} {\bibfnamefont {M.}~\bibnamefont {Vojta}},\ }\bibinfo {title}
  {Metastable superfluidity of repulsive fermionic atoms in optical lattices},\
  \href {\doibase 10.1103/PhysRevLett.101.265301} {\bibfield  {journal}
  {\bibinfo  {journal} {Phys. Rev. Lett.}\ }\textbf {\bibinfo {volume} {101}},\
  \bibinfo {pages} {265301} (\bibinfo {year} {2008})}\BibitemShut {NoStop}%
\bibitem [{\citenamefont {Shimano}\ and\ \citenamefont
  {Tsuji}(2020)}]{annurev:/content/journals/10.1146/annurev-conmatphys-031119-050813}%
  \BibitemOpen
  \bibfield  {author} {\bibinfo {author} {\bibfnamefont {R.}~\bibnamefont
  {Shimano}}\ and\ \bibinfo {author} {\bibfnamefont {N.}~\bibnamefont
  {Tsuji}},\ }\bibinfo {title} {Higgs mode in superconductors},\ \href
  {\doibase https://doi.org/10.1146/annurev-conmatphys-031119-050813}
  {\bibfield  {journal} {\bibinfo  {journal} {Annu. Rev. Condens. Matter
  Phys.}\ }\textbf {\bibinfo {volume} {11}},\ \bibinfo {pages} {103} (\bibinfo
  {year} {2020})}\BibitemShut {NoStop}%
\bibitem [{\citenamefont {Vidal}(2007)}]{iTEBD}%
  \BibitemOpen
  \bibfield  {author} {\bibinfo {author} {\bibfnamefont {G.}~\bibnamefont
  {Vidal}},\ }\bibinfo {title} {Classical simulation of infinite-size quantum
  lattice systems in one spatial dimension},\ \href {\doibase
  10.1103/PhysRevLett.98.070201} {\bibfield  {journal} {\bibinfo  {journal}
  {Phys. Rev. Lett.}\ }\textbf {\bibinfo {volume} {98}},\ \bibinfo {pages}
  {070201} (\bibinfo {year} {2007})}\BibitemShut {NoStop}%
\bibitem [{\citenamefont {{Supplemental Material}}(2025)}]{SM}%
  \BibitemOpen
  \bibinfo {note} {See \hyperref[sec:SM]{Supplemental Material} for details.}
  \BibitemShut {Stop}%
\bibitem [{\citenamefont {Ejima}\ \emph
  {et~al.}(2020{\natexlab{a}})\citenamefont {Ejima}, \citenamefont {Kaneko},
  \citenamefont {Lange}, \citenamefont {Yunoki},\ and\ \citenamefont
  {Fehske}}]{SCES19}%
  \BibitemOpen
  \bibfield  {author} {\bibinfo {author} {\bibfnamefont {S.}~\bibnamefont
  {Ejima}}, \bibinfo {author} {\bibfnamefont {T.}~\bibnamefont {Kaneko}},
  \bibinfo {author} {\bibfnamefont {F.}~\bibnamefont {Lange}}, \bibinfo
  {author} {\bibfnamefont {S.}~\bibnamefont {Yunoki}}, \ and\ \bibinfo {author}
  {\bibfnamefont {H.}~\bibnamefont {Fehske}},\ }\bibinfo {title} {Photoinduced
  $\eta$-pairing in one-dimensional Mott insulators},\ \href {\doibase
  10.7566/JPSCP.30.011184} {\bibfield  {journal} {\bibinfo  {journal} {JPS
  Conf. Proc.}\ }\textbf {\bibinfo {volume} {30}},\ \bibinfo {pages} {011184}
  (\bibinfo {year} {2020}{\natexlab{a}})}\BibitemShut {NoStop}%
\bibitem [{\citenamefont {Shao}\ \emph {et~al.}(2016)\citenamefont {Shao},
  \citenamefont {Tohyama}, \citenamefont {Luo},\ and\ \citenamefont
  {Lu}}]{PhysRevB.93.195144}%
  \BibitemOpen
  \bibfield  {author} {\bibinfo {author} {\bibfnamefont {C.}~\bibnamefont
  {Shao}}, \bibinfo {author} {\bibfnamefont {T.}~\bibnamefont {Tohyama}},
  \bibinfo {author} {\bibfnamefont {H.-G.}\ \bibnamefont {Luo}}, \ and\
  \bibinfo {author} {\bibfnamefont {H.}~\bibnamefont {Lu}},\ }\bibinfo {title}
  {Numerical method to compute optical conductivity based on pump-probe
  simulations},\ \href {\doibase 10.1103/PhysRevB.93.195144} {\bibfield
  {journal} {\bibinfo  {journal} {Phys. Rev. B}\ }\textbf {\bibinfo {volume}
  {93}},\ \bibinfo {pages} {195144} (\bibinfo {year} {2016})}\BibitemShut
  {NoStop}%
\bibitem [{\citenamefont {Rinc\'on}\ and\ \citenamefont
  {Feiguin}(2021)}]{PhysRevB.104.085122}%
  \BibitemOpen
  \bibfield  {author} {\bibinfo {author} {\bibfnamefont {J.}~\bibnamefont
  {Rinc\'on}}\ and\ \bibinfo {author} {\bibfnamefont {A.~E.}\ \bibnamefont
  {Feiguin}},\ }\bibinfo {title} {Nonequilibrium optical response of a
  one-dimensional Mott insulator},\ \href {\doibase
  10.1103/PhysRevB.104.085122} {\bibfield  {journal} {\bibinfo  {journal}
  {Phys. Rev. B}\ }\textbf {\bibinfo {volume} {104}},\ \bibinfo {pages}
  {085122} (\bibinfo {year} {2021})}\BibitemShut {NoStop}%
\bibitem [{\citenamefont {Sugimoto}\ and\ \citenamefont
  {Ejima}(2023)}]{Sugimoto23}%
  \BibitemOpen
  \bibfield  {author} {\bibinfo {author} {\bibfnamefont {K.}~\bibnamefont
  {Sugimoto}}\ and\ \bibinfo {author} {\bibfnamefont {S.}~\bibnamefont
  {Ejima}},\ }\bibinfo {title} {Pump-probe spectroscopy of the one-dimensional
  extended Hubbard model at half filling},\ \href {\doibase
  10.1103/PhysRevB.108.195128} {\bibfield  {journal} {\bibinfo  {journal}
  {Phys. Rev. B}\ }\textbf {\bibinfo {volume} {108}},\ \bibinfo {pages}
  {195128} (\bibinfo {year} {2023})}\BibitemShut {NoStop}%
\bibitem [{\citenamefont {Li}\ \emph {et~al.}(2020)\citenamefont {Li},
  \citenamefont {Golez}, \citenamefont {Werner},\ and\ \citenamefont
  {Eckstein}}]{PhysRevB.102.165136}%
  \BibitemOpen
  \bibfield  {author} {\bibinfo {author} {\bibfnamefont {J.}~\bibnamefont
  {Li}}, \bibinfo {author} {\bibfnamefont {D.}~\bibnamefont {Golez}}, \bibinfo
  {author} {\bibfnamefont {P.}~\bibnamefont {Werner}}, \ and\ \bibinfo {author}
  {\bibfnamefont {M.}~\bibnamefont {Eckstein}},\ }\bibinfo {title}
  {$\ensuremath{\eta}$-paired superconducting hidden phase in photodoped Mott
  insulators},\ \href {\doibase 10.1103/PhysRevB.102.165136} {\bibfield
  {journal} {\bibinfo  {journal} {Phys. Rev. B}\ }\textbf {\bibinfo {volume}
  {102}},\ \bibinfo {pages} {165136} (\bibinfo {year} {2020})}\BibitemShut
  {NoStop}%
\bibitem [{\citenamefont {Ray}\ \emph {et~al.}(2023)\citenamefont {Ray},
  \citenamefont {Murakami},\ and\ \citenamefont
  {Werner}}]{PhysRevB.108.174515}%
  \BibitemOpen
  \bibfield  {author} {\bibinfo {author} {\bibfnamefont {S.}~\bibnamefont
  {Ray}}, \bibinfo {author} {\bibfnamefont {Y.}~\bibnamefont {Murakami}}, \
  and\ \bibinfo {author} {\bibfnamefont {P.}~\bibnamefont {Werner}},\ }\bibinfo
  {title} {Nonthermal superconductivity in photodoped multiorbital Hubbard
  systems},\ \href {\doibase 10.1103/PhysRevB.108.174515} {\bibfield  {journal}
  {\bibinfo  {journal} {Phys. Rev. B}\ }\textbf {\bibinfo {volume} {108}},\
  \bibinfo {pages} {174515} (\bibinfo {year} {2023})}\BibitemShut {NoStop}%
\bibitem [{\citenamefont {Park}\ and\ \citenamefont
  {Light}(1986)}]{JChemPhys86}%
  \BibitemOpen
  \bibfield  {author} {\bibinfo {author} {\bibfnamefont {T.~J.}\ \bibnamefont
  {Park}}\ and\ \bibinfo {author} {\bibfnamefont {J.~C.}\ \bibnamefont
  {Light}},\ }\bibinfo {title} {{Unitary quantum time evolution by iterative
  Lanczos reduction}},\ \href {\doibase 10.1063/1.451548} {\bibfield  {journal}
  {\bibinfo  {journal} {J. Chem. Phys.}\ }\textbf {\bibinfo {volume} {85}},\
  \bibinfo {pages} {5870} (\bibinfo {year} {1986})}\BibitemShut {NoStop}%
\bibitem [{\citenamefont {Ejima}\ \emph
  {et~al.}(2020{\natexlab{b}})\citenamefont {Ejima}, \citenamefont {Kaneko},
  \citenamefont {Lange}, \citenamefont {Yunoki},\ and\ \citenamefont
  {Fehske}}]{Ejima20}%
  \BibitemOpen
  \bibfield  {author} {\bibinfo {author} {\bibfnamefont {S.}~\bibnamefont
  {Ejima}}, \bibinfo {author} {\bibfnamefont {T.}~\bibnamefont {Kaneko}},
  \bibinfo {author} {\bibfnamefont {F.}~\bibnamefont {Lange}}, \bibinfo
  {author} {\bibfnamefont {S.}~\bibnamefont {Yunoki}}, \ and\ \bibinfo {author}
  {\bibfnamefont {H.}~\bibnamefont {Fehske}},\ }\bibinfo {title} {Photoinduced
  $\ensuremath{\eta}$-pairing at finite temperatures},\ \href {\doibase
  10.1103/PhysRevResearch.2.032008} {\bibfield  {journal} {\bibinfo  {journal}
  {Phys. Rev. Research}\ }\textbf {\bibinfo {volume} {2}},\ \bibinfo {pages}
  {032008} (\bibinfo {year} {2020}{\natexlab{b}})}\BibitemShut {NoStop}%
\bibitem [{\citenamefont {Higgs}(1964)}]{PhysRevLett.13.508}%
  \BibitemOpen
  \bibfield  {author} {\bibinfo {author} {\bibfnamefont {P.~W.}\ \bibnamefont
  {Higgs}},\ }\bibinfo {title} {Broken symmetries and the masses of gauge
  bosons},\ \href {\doibase 10.1103/PhysRevLett.13.508} {\bibfield  {journal}
  {\bibinfo  {journal} {Phys. Rev. Lett.}\ }\textbf {\bibinfo {volume} {13}},\
  \bibinfo {pages} {508} (\bibinfo {year} {1964})}\BibitemShut {NoStop}%
\bibitem [{\citenamefont {Varma}(2002)}]{Varma2002}%
  \BibitemOpen
  \bibfield  {author} {\bibinfo {author} {\bibfnamefont {C.~M.}\ \bibnamefont
  {Varma}},\ }\bibinfo {title} {Higgs boson in superconductors},\ \href
  {\doibase 10.1023/A:1013890507658} {\bibfield  {journal} {\bibinfo  {journal}
  {J. Low Temp. Phys.}\ }\textbf {\bibinfo {volume} {126}},\ \bibinfo {pages}
  {901} (\bibinfo {year} {2002})}\BibitemShut {NoStop}%
\bibitem [{\citenamefont {Matsunaga}\ and\ \citenamefont
  {Shimano}(2012)}]{PhysRevLett.109.187002}%
  \BibitemOpen
  \bibfield  {author} {\bibinfo {author} {\bibfnamefont {R.}~\bibnamefont
  {Matsunaga}}\ and\ \bibinfo {author} {\bibfnamefont {R.}~\bibnamefont
  {Shimano}},\ }\bibinfo {title} {Nonequilibrium bcs state dynamics induced by
  intense terahertz pulses in a superconducting nbn film},\ \href {\doibase
  10.1103/PhysRevLett.109.187002} {\bibfield  {journal} {\bibinfo  {journal}
  {Phys. Rev. Lett.}\ }\textbf {\bibinfo {volume} {109}},\ \bibinfo {pages}
  {187002} (\bibinfo {year} {2012})}\BibitemShut {NoStop}%
\bibitem [{\citenamefont {Matsunaga}\ \emph {et~al.}(2013)\citenamefont
  {Matsunaga}, \citenamefont {Hamada}, \citenamefont {Makise}, \citenamefont
  {Uzawa}, \citenamefont {Terai}, \citenamefont {Wang},\ and\ \citenamefont
  {Shimano}}]{PhysRevLett.111.057002}%
  \BibitemOpen
  \bibfield  {author} {\bibinfo {author} {\bibfnamefont {R.}~\bibnamefont
  {Matsunaga}}, \bibinfo {author} {\bibfnamefont {Y.~I.}\ \bibnamefont
  {Hamada}}, \bibinfo {author} {\bibfnamefont {K.}~\bibnamefont {Makise}},
  \bibinfo {author} {\bibfnamefont {Y.}~\bibnamefont {Uzawa}}, \bibinfo
  {author} {\bibfnamefont {H.}~\bibnamefont {Terai}}, \bibinfo {author}
  {\bibfnamefont {Z.}~\bibnamefont {Wang}}, \ and\ \bibinfo {author}
  {\bibfnamefont {R.}~\bibnamefont {Shimano}},\ }\bibinfo {title} {Higgs
  amplitude mode in the bcs superconductors
  ${\mathrm{Nb}}_{1\mathrm{\text{\ensuremath{-}}}x}{\mathrm{Ti}}_{x}\mathbf{N}$
  induced by terahertz pulse excitation},\ \href {\doibase 10.1103/PhysRevLett.111.057002} {\bibfield  {journal} {\bibinfo  {journal}
  {Phys. Rev. Lett.}\ }\textbf {\bibinfo {volume} {111}},\ \bibinfo {pages}
  {057002} (\bibinfo {year} {2013})}\BibitemShut {NoStop}%
\bibitem [{\citenamefont {Schwarz}\ \emph
  {et~al.}(2020{\natexlab{a}})\citenamefont {Schwarz}, \citenamefont
  {Fauseweh}, \citenamefont {Tsuji}, \citenamefont {Cheng}, \citenamefont
  {Bittner}, \citenamefont {Krull}, \citenamefont {Berciu}, \citenamefont
  {Uhrig}, \citenamefont {Schnyder}, \citenamefont {Kaiser},\ and\
  \citenamefont {Manske}}]{Schwarz2020}%
  \BibitemOpen
  \bibfield  {author} {\bibinfo {author} {\bibfnamefont {L.}~\bibnamefont
  {Schwarz}}, \bibinfo {author} {\bibfnamefont {B.}~\bibnamefont {Fauseweh}},
  \bibinfo {author} {\bibfnamefont {N.}~\bibnamefont {Tsuji}}, \bibinfo
  {author} {\bibfnamefont {N.}~\bibnamefont {Cheng}}, \bibinfo {author}
  {\bibfnamefont {N.}~\bibnamefont {Bittner}}, \bibinfo {author} {\bibfnamefont
  {H.}~\bibnamefont {Krull}}, \bibinfo {author} {\bibfnamefont
  {M.}~\bibnamefont {Berciu}}, \bibinfo {author} {\bibfnamefont {G.~S.}\
  \bibnamefont {Uhrig}}, \bibinfo {author} {\bibfnamefont {A.~P.}\ \bibnamefont
  {Schnyder}}, \bibinfo {author} {\bibfnamefont {S.}~\bibnamefont {Kaiser}}, \
  and\ \bibinfo {author} {\bibfnamefont {D.}~\bibnamefont {Manske}},\ }\bibinfo
  {title} {Classification and characterization of nonequilibrium Higgs modes in
  unconventional superconductors},\ \href {\doibase 10.1038/s41467-019-13763-5}
  {\bibfield  {journal} {\bibinfo  {journal} {Nat. Commun.}\ }\textbf {\bibinfo
  {volume} {11}},\ \bibinfo {pages} {287} (\bibinfo {year}
  {2020}{\natexlab{a}})}\BibitemShut {NoStop}%
\bibitem [{\citenamefont {Schwarz}\ \emph
  {et~al.}(2020{\natexlab{b}})\citenamefont {Schwarz}, \citenamefont
  {Fauseweh},\ and\ \citenamefont {Manske}}]{PhysRevB.101.224510}%
  \BibitemOpen
  \bibfield  {author} {\bibinfo {author} {\bibfnamefont {L.}~\bibnamefont
  {Schwarz}}, \bibinfo {author} {\bibfnamefont {B.}~\bibnamefont {Fauseweh}}, \
  and\ \bibinfo {author} {\bibfnamefont {D.}~\bibnamefont {Manske}},\ }\bibinfo
  {title} {Momentum-resolved analysis of condensate dynamic and Higgs
  oscillations in quenched superconductors with time-resolved arpes},\ \href
  {\doibase 10.1103/PhysRevB.101.224510} {\bibfield  {journal} {\bibinfo
  {journal} {Phys. Rev. B}\ }\textbf {\bibinfo {volume} {101}},\ \bibinfo
  {pages} {224510} (\bibinfo {year} {2020}{\natexlab{b}})}\BibitemShut
  {NoStop}%
\bibitem [{\citenamefont {Ido}\ \emph {et~al.}(2017)\citenamefont {Ido},
  \citenamefont {Ohgoe},\ and\ \citenamefont
  {Imada}}]{doi:10.1126/sciadv.1700718}%
  \BibitemOpen
  \bibfield  {author} {\bibinfo {author} {\bibfnamefont {K.}~\bibnamefont
  {Ido}}, \bibinfo {author} {\bibfnamefont {T.}~\bibnamefont {Ohgoe}}, \ and\
  \bibinfo {author} {\bibfnamefont {M.}~\bibnamefont {Imada}},\ }\bibinfo
  {title} {Correlation-induced superconductivity dynamically stabilized and
  enhanced by laser irradiation},\ \href {\doibase 10.1126/sciadv.1700718}
  {\bibfield  {journal} {\bibinfo  {journal} {Sci. Adv.}\ }\textbf {\bibinfo
  {volume} {3}},\ \bibinfo {pages} {e1700718} (\bibinfo {year}
  {2017})}\BibitemShut {NoStop}%
\bibitem [{\citenamefont {Buzzi}\ \emph {et~al.}(2021)\citenamefont {Buzzi},
  \citenamefont {Jotzu}, \citenamefont {Cavalleri}, \citenamefont {Cirac},
  \citenamefont {Demler}, \citenamefont {Halperin}, \citenamefont {Lukin},
  \citenamefont {Shi}, \citenamefont {Wang},\ and\ \citenamefont
  {Podolsky}}]{PhysRevX.11.011055}%
  \BibitemOpen
  \bibfield  {author} {\bibinfo {author} {\bibfnamefont {M.}~\bibnamefont
  {Buzzi}}, \bibinfo {author} {\bibfnamefont {G.}~\bibnamefont {Jotzu}},
  \bibinfo {author} {\bibfnamefont {A.}~\bibnamefont {Cavalleri}}, \bibinfo
  {author} {\bibfnamefont {J.~I.}\ \bibnamefont {Cirac}}, \bibinfo {author}
  {\bibfnamefont {E.~A.}\ \bibnamefont {Demler}}, \bibinfo {author}
  {\bibfnamefont {B.~I.}\ \bibnamefont {Halperin}}, \bibinfo {author}
  {\bibfnamefont {M.~D.}\ \bibnamefont {Lukin}}, \bibinfo {author}
  {\bibfnamefont {T.}~\bibnamefont {Shi}}, \bibinfo {author} {\bibfnamefont
  {Y.}~\bibnamefont {Wang}}, \ and\ \bibinfo {author} {\bibfnamefont
  {D.}~\bibnamefont {Podolsky}},\ }\bibinfo {title} {Higgs-mediated optical
  amplification in a nonequilibrium superconductor},\ \href {\doibase
  10.1103/PhysRevX.11.011055} {\bibfield  {journal} {\bibinfo  {journal} {Phys.
  Rev. X}\ }\textbf {\bibinfo {volume} {11}},\ \bibinfo {pages} {011055}
  (\bibinfo {year} {2021})}\BibitemShut {NoStop}%
\bibitem [{\citenamefont {Fishman}\ \emph {et~al.}(2022)\citenamefont
  {Fishman}, \citenamefont {White},\ and\ \citenamefont
  {Stoudenmire}}]{ITensor}%
  \BibitemOpen
  \bibfield  {author} {\bibinfo {author} {\bibfnamefont {M.}~\bibnamefont
  {Fishman}}, \bibinfo {author} {\bibfnamefont {S.~R.}\ \bibnamefont {White}},\
  and\ \bibinfo {author} {\bibfnamefont {E.~M.}\ \bibnamefont {Stoudenmire}},\
  }\bibfield  {title} {\bibinfo {title}\
  {The ITensor Software Library for Tensor Network Calculations},}\
  \href {https://doi.org/10.21468/SciPostPhysCodeb.4} 
  {\bibfield  {journal}
  {\bibinfo  {journal} {SciPost Phys. Codebases}\ \bibinfo {pages} {4}}
  (\bibinfo {year} {2022})}.
  \BibitemShut {Stop}%
\label{LastBibItem}
\end{thebibliography}

\begin{thebibliography}{4}%
\makeatletter
\providecommand \@ifxundefined [1]{%
 \@ifx{#1\undefined}
}%
\providecommand \@ifnum [1]{%
 \ifnum #1\expandafter \@firstoftwo
 \else \expandafter \@secondoftwo
 \fi
}%
\providecommand \@ifx [1]{%
 \ifx #1\expandafter \@firstoftwo
 \else \expandafter \@secondoftwo
 \fi
}%
\providecommand \natexlab [1]{#1}%
\providecommand \enquote  [1]{``#1''}%
\providecommand \bibnamefont  [1]{#1}%
\providecommand \bibfnamefont [1]{#1}%
\providecommand \citenamefont [1]{#1}%
\providecommand \href@noop [0]{\@secondoftwo}%
\providecommand \href [0]{\begingroup \@sanitize@url \@href}%
\providecommand \@href[1]{\@@startlink{#1}\@@href}%
\providecommand \@@href[1]{\endgroup#1\@@endlink}%
\providecommand \@sanitize@url [0]{\catcode `\\12\catcode `\$12\catcode
  `\&12\catcode `\#12\catcode `\^12\catcode `\_12\catcode `\%12\relax}%
\providecommand \@@startlink[1]{}%
\providecommand \@@endlink[0]{}%
\providecommand \url  [0]{\begingroup\@sanitize@url \@url }%
\providecommand \@url [1]{\endgroup\@href {#1}{\urlprefix }}%
\providecommand \urlprefix  [0]{URL }%
\providecommand \Eprint [0]{\href }%
\providecommand \doibase [0]{https://doi.org/}%
\providecommand \selectlanguage [0]{\@gobble}%
\providecommand \bibinfo  [0]{\@secondoftwo}%
\providecommand \bibfield  [0]{\@secondoftwo}%
\providecommand \translation [1]{[#1]}%
\providecommand \BibitemOpen [0]{}%
\providecommand \bibitemStop [0]{}%
\providecommand \bibitemNoStop [0]{.\EOS\space}%
\providecommand \EOS [0]{\spacefactor3000\relax}%
\providecommand \BibitemShut  [1]{\csname bibitem#1\endcsname}%
\let\auto@bib@innerbib\@empty
\bibitem [{\citenamefont {Vidal}(2007)}]{SM-iTEBD}%
  \BibitemOpen
  \bibfield  {author} {\bibinfo {author} {\bibfnamefont {G.}~\bibnamefont
  {Vidal}},\ }\href {https://doi.org/10.1103/PhysRevLett.98.070201} {\bibfield
  {journal} {\bibinfo  {journal} {Phys. Rev. Lett.}\ }\textbf {\bibinfo
  {volume} {98}},\ \bibinfo {pages} {070201} (\bibinfo {year}
  {2007})}\BibitemShut {NoStop}%
\bibitem [{\citenamefont {Sugimoto}\ and\ \citenamefont
  {Ejima}(2023)}]{SM-Sugimoto23}%
  \BibitemOpen
  \bibfield  {author} {\bibinfo {author} {\bibfnamefont {K.}~\bibnamefont
  {Sugimoto}}\ and\ \bibinfo {author} {\bibfnamefont {S.}~\bibnamefont
  {Ejima}},\ }\href {https://doi.org/10.1103/PhysRevB.108.195128} {\bibfield
  {journal} {\bibinfo  {journal} {Phys. Rev. B}\ }\textbf {\bibinfo {volume}
  {108}},\ \bibinfo {pages} {195128} (\bibinfo {year} {2023})}\BibitemShut
  {NoStop}%
\bibitem [{\citenamefont {Li}\ \emph {et~al.}(2020)\citenamefont {Li},
  \citenamefont {Golez}, \citenamefont {Werner},\ and\ \citenamefont
  {Eckstein}}]{SM-PhysRevB.102.165136}%
  \BibitemOpen
  \bibfield  {author} {\bibinfo {author} {\bibfnamefont {J.}~\bibnamefont
  {Li}}, \bibinfo {author} {\bibfnamefont {D.}~\bibnamefont {Golez}}, \bibinfo
  {author} {\bibfnamefont {P.}~\bibnamefont {Werner}},\ and\ \bibinfo {author}
  {\bibfnamefont {M.}~\bibnamefont {Eckstein}},\ }\href
  {https://doi.org/10.1103/PhysRevB.102.165136} {\bibfield  {journal} {\bibinfo
   {journal} {Phys. Rev. B}\ }\textbf {\bibinfo {volume} {102}},\ \bibinfo
  {pages} {165136} (\bibinfo {year} {2020})}\BibitemShut {NoStop}%
\bibitem [{\citenamefont {Ray}\ \emph {et~al.}(2023)\citenamefont {Ray},
  \citenamefont {Murakami},\ and\ \citenamefont
  {Werner}}]{SM-PhysRevB.108.174515}%
  \BibitemOpen
  \bibfield  {author} {\bibinfo {author} {\bibfnamefont {S.}~\bibnamefont
  {Ray}}, \bibinfo {author} {\bibfnamefont {Y.}~\bibnamefont {Murakami}},\ and\
  \bibinfo {author} {\bibfnamefont {P.}~\bibnamefont {Werner}},\ }\href
  {https://doi.org/10.1103/PhysRevB.108.174515} {\bibfield  {journal} {\bibinfo
   {journal} {Phys. Rev. B}\ }\textbf {\bibinfo {volume} {108}},\ \bibinfo
  {pages} {174515} (\bibinfo {year} {2023})}\BibitemShut {NoStop}%
\end{thebibliography}
%

\clearpage

\appendix
\renewcommand\thesection{}
\renewcommand{\theequation}{S\arabic{equation}}
\setcounter{equation}{0}
\renewcommand\thefigure{S.\arabic{figure}}
\setcounter{figure}{0}
\renewcommand{\bibnumfmt}[1]{[S#1]}
\renewcommand{\citenumfont}[1]{S#1}

\onecolumngrid
\phantomsection
\label{sec:SM}

\widetext
\begin{center}
 \textbf{\large Supplemental Material: Radiative Higgs mode in photoinduced $\eta$ pairs}
\end{center}

\begin{center}
 Satoshi Ejima$^{1,2}$ and Benedikt Fauseweh$^{3,4}$\\[5pt]
 \textit{
 $^1$ Institute of Software Technology, German Aerospace Center (DLR), 22529 Hamburg, Germany\\
 $^2$ Computational Condensed Matter Physics Laboratory,
     RIKEN Cluster for Pioneering Research (CPR), Saitama 351-0198, Japan\\
 $^3$ Institute of Software Technology, German Aerospace Center (DLR), 51147 Cologne, Germany\\
 $^4$ Department of Physics, TU Dortmund University, Otto-Hahn-Str.~4, 44227 Dortmund, Germany
 }
 \end{center}

\vspace*{5pt}


\section{Choice of pump-pulse parameters}
In Fig. 1 of the main text, we selected three representative points in the parameter space of the amplitude $A_0$ and the frequency $\omega_{\rm p}$ for the fixed width $\sigma_{\rm p}=2$ centered at time $t_0=10\tH^{-1}$ to illustrate the different dynamical regimes observed in our study. These points were chosen based on the following considerations:

\begin{itemize}
 \item Point + ($A_0=0.4$ and $\omega_{\rm p}/\tH=7.0$): the $\eta$-pairing dominated point, where the pair correlations $\tilde{P}(q=\pi,t)$ is most prominent.
 \item Point $\times$ ($A_0=0.4$ and $\omega_{\rm p}/\tH=4.0$): the off-resonant point where neither coherent pairing nor significant doublon formation occurs.
 \item Point $\diamond$ ($A_0=0.95$ and $\omega_{\rm p}/\tH=8.4$): the photoinduced doublon dominated point where the doublon occupancy $2n_{\rm d}(t)$ is most significant. 
\end{itemize}

These choices allow us to highlight the qualitative differences in the real-time optical conductivity and pairing correlations across different excitation conditions.

\section{Numerical technique}

\begin{figure}[htb]
 \includegraphics[width=0.5\linewidth]{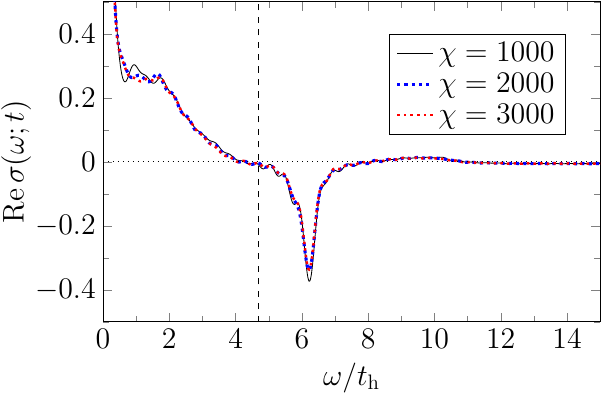}
 \caption{Bond-dimension dependence of the nonequilibrium optical conductivity $\rm{Re}\,\sigma(\omega;t=25\tH^{-1})$ with the pump pulse parameters, $A_0=0.4$, $\omega_{\rm p}=7.0$ and $\sigma_{\rm p}=2.0$ centered at time $t_0=10\tH^{-1}$. The probe pulse parameters are $A_0^{\rm pr}=0.05$ and $\sigma_{\rm pr}=0.05$.
 The vertical dashed line represents the position of the Mott gap $\omega = \Delta_{\rm c}$.
 }
 \phantomsection
 \label{fig:supp:chi-dependence}
\end{figure}

In this study, we utilize the infinite time-evolved block decimation (iTEBD) technique~\citeSM{SM-iTEBD} to compute the optical conductivity in nonequilibrium systems, ${\rm Re}\,\sigma(\omega; t)$, defined in Eq.~(7) in the main text. See also Ref.~\citeSM{SM-Sugimoto23}. The real-time evolution was performed using the second-order Suzuki-Trotter decomposition. Figure~\ref{fig:supp:chi-dependence} presents the bond-dimension dependence of ${\rm Re}\,\sigma(\omega; t)$. 
As illustrated in the figure, the bond-dimension dependence is negligible in the region of $\omega > \Delta_{\rm c}$. While minor differences are observed between $\chi=1000$ and $\chi=2000$, the results exhibit clear convergence upon increasing the bond dimension to $\chi=3000$, which is employed for the calculation presented in Fig.~2. During the simulation for Fig.~2, the discarded weights remain lower than $1\times10^{-6}$ with $\chi=3000$.

\section{Imaginary part of the nonequilibrium optical conductivity}
\begin{figure}[t]
 \includegraphics[width=0.99\linewidth]{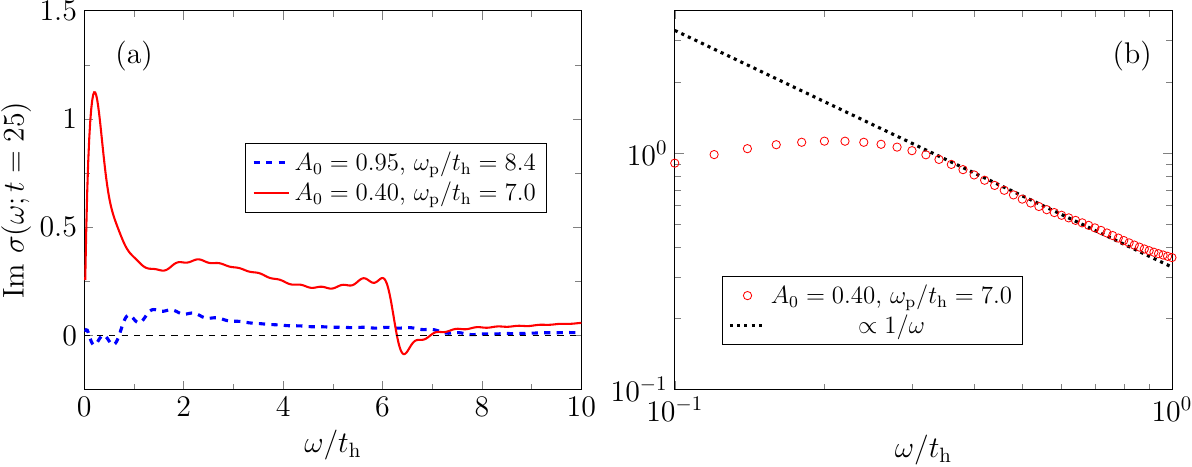}
 \caption{(a): Imaginary part of the nonequilibrium optical conductivity in the Hubbard model with $U/\tH=8$ at the $\eta$-pairing-dominant (denoted by the '+' symbol in Fig.~1 with $A_0=0.4$ and $\omega_{\rm p}=7.0$) and doublon-dominant (denoted by the '$\diamond$' symbol in Fig.~1 with $A_0=0.95$ and $\omega_{\rm p}=8.4$) points. The other pump-probe parameters are $\sigma_{\rm p}=2.0$ centered at time $t_0=10\tH^{-1}$. The probe pulse parameters are $A_0^{\rm pr}=0.05$ and $\sigma_{\rm pr}=0.05$.
    (b): Log-log plot of ${\rm Im}\,\sigma(\omega; t)$ at low frequency in the $\eta$-pairing-dominant regime (symbols) scaled by $1/\omega$.
 }
 \phantomsection
 \label{fig:supp:im-sw}
\end{figure}

In the main text, we have investigated only the real part of the nonequilibrium optical conductivity ${\rm Re}\,\sigma(\omega; t)$ in order to demonstrate the characteristic negative spectral weight at the $\eta$-pairing-dominant point.
As shown in Refs.~\citeSM{SM-PhysRevB.102.165136,SM-PhysRevB.108.174515}, the imaginary part of the optical conductivity ${\rm Im}\,\sigma(\omega; t)$ also provides us with the signature of the superconducting character as a $1/\omega$ scaling in the low frequency region.

In this section, we present the numerical results of ${\rm Im}\,\sigma(\omega; t)$ in the $\eta$-pairing-dominant regime, to investigate the possibility of scaling $1/\omega$ in the limit of $\omega\to 0^+$.
As shown in Fig.~\ref{fig:supp:im-sw}(a), the spectral weights exhibit a significant increase in the low-frequency limit, while remaining very small in the doublon-dominant region. Furthermore, the log-log plot [Fig.~\ref{fig:supp:im-sw}(b)] confirms that the data in the range $0.1\lesssim \omega \lesssim 1.0$ can be reasonably scaled by $1/\omega$. 

Note that a proper analysis of the $\omega\to0^+$  behavior requires longer-time simulations than those currently performed. Therefore, a more detailed investigation of this issue is left for future work.

%

\end{document}